\documentclass[twocolumn]{aastex631}

\usepackage{xspace}
\usepackage{amsmath}
\usepackage{url}
\usepackage{listings}
\usepackage{verbatim}
\usepackage[normalem]{ulem}
\usepackage{lipsum, babel, ulem, soul, cancel}
\usepackage[shortlabels]{enumitem}

\newcommand{\cosmic}{{\small COSMIC}\xspace}

\newcommand{\galah}{{\small GALAH}\xspace}
\newcommand{\apogee}{{\small APOGEE}\xspace}

\newcommand{\lamost}{{\small LAMOST}\xspace}

\newcommand{\gaia}{\textit{Gaia}\xspace}

\newcommand{\solmass}{M$_\odot$\xspace}
\newcommand{\solrad}{R$_\odot$\xspace}
\newcommand{\app}{$\sim$}
\newcommand{\logg}{$\log g$\xspace}
\newcommand{\teff}{T$_\textrm{eff}$\xspace}

\newcommand{\ali}{$\rm A(Li)$\xspace}

\newcommand{\kms}{km s$^{-1}$\xspace}

\newcommand{\ang}{$\textrm{\AA}$\xspace}

\newcommand{\li}{lithium\xspace}
\newcommand{\lirich}{Li--rich\xspace}
\newcommand{\lin}{Li--normal\xspace}
\newcommand{\dg}{doppelg\"anger\xspace}
\newcommand{\dgs}{doppelg\"angers\xspace}
\newcommand{\mt}{mass transfer\xspace}
\newcommand{\ce}{common envelope\xspace}
\newcommand{\cheb}{core helium--burning\xspace}

\newcommand{\sprocess}{$s-$process\xspace}

\newcommand{\columbia}{Department of Astronomy, Columbia University, 550 West 120th Street, New York, NY, USA}
\newcommand{\anu}{Research School of Astronomy \& Astrophysics, Australian National
University, Canberra, ACT 2611, Australia}
\newcommand{\cca}{Center for Computational Astrophysics, Flatiron Institute, 162 5th Avenue, Manhattan, NY, USA}
\newcommand{\arc}{ARC Centre of Excellence for All Sky Astrophysics in 3 Dimensions (ASTRO 3D)}
\newcommand{\monash}{School of Physics \& Astronomy, Monash University, Clayton 3800, Victoria, Australia}

\shorttitle{Binary Architectures of Lithium--Rich Giants}
\shortauthors{Sayeed et al.}

\begin{document}

\title{Probing Binary Architectures of Lithium--Rich Giants in \galah with \cosmic and Stellar Models}

\correspondingauthor{Maryum Sayeed}
\email{maryum.sayeed@columbia.edu}

\author[0000-0001-6180-8482]{Maryum Sayeed}
\affiliation{\columbia}

\author[0009-0002-7777-2351]{Selina Yang}
\affiliation{\columbia}

\author[0000-0001-7902-8134]{Giulia Cinquegrana}
\affiliation{\monash}
\affiliation{\arc}

\author[0000-0001-5082-6693]{Melissa K. Ness}
\affiliation{\columbia}
\affiliation{\anu}

\author[0000-0001-5228-6598]{Katelyn Breivik}
\affiliation{McWilliams Center for Cosmology and Astrophysics, Department of Physics, Carnegie Mellon University, Pittsburgh, PA 15213, USA}
\affiliation{\cca}

\author[0000-0003-0174-0564]{Andrew R. Casey}
\affiliation{\monash}
\affiliation{\arc}
\affiliation{\cca}

\author[0000-0002-4031-8553]{Sven Buder}
\affiliation{\arc}
\affiliation{\anu}

\author[0000-0002-3625-6951]{Amanda I. Karakas}
\affiliation{\monash}
\affiliation{\arc}

\begin{abstract}
Surface lithium is depleted when a star goes through the first dredge--up phase, yet 1\% of red giants are found to be Li--rich. The formation mechanism for these remains uncertain. We combine observational constraints from \galah \lirich giants, with the binary population synthesis code \cosmic to investigate system properties of these objects assuming binary mass transfer. By evolving 9 million binary systems, we find that binary histories most consistent with observational constraints are mass transfer from an intermediate--mass AGB donor to a main--sequence star now observed as a \lirich red giant. In \galah, 9\% of main--sequence stars have $\rm A(Li)=2.5-3.2$ dex making it plausible to create red giants with $\rm A(Li)=1.5-2.2 \; dex$ via main--sequence mass transfer, but cannot explain the more enriched giants $\rm A(Li) \gtrsim 2.2 \; dex$. Nucleosynthetic yields from stellar models show that AGB stars with initial masses of $4.25-5 \; \rm M_\odot$ and $8 \; \rm M_\odot$ contain the most Li in their ejecta. Intermediate--mass AGB stars comprise 29\% of COSMIC results, with present--day separations $s=3.3\pm0.5 \rm \; AU$ and mass ratios $q=0.5-1.6$. We achieve 95\% agreement in mean enhancements in $\rm (Ba, Y)$ between GALAH observations and stellar models of 6 and $8 \rm \; M_\odot$ AGB, assuming 1\% mass transfer efficiency. We find a low mass transfer efficiency best reproduces GALAH observations suggesting that the preferred mass transfer mechanism for Li--enrichment is via wind Roche Lobe Overflow. While we constrain the most plausible binary parameters assuming AGB mass transfer creates \lirich giants, discrepancies in nucleosynthesis comparisons, and the small fraction of Li--enhanced main--sequence stars suggests additional enrichment mechanisms are likely. 
\end{abstract}

\keywords{Binary stars (154) -- chemically peculiar stars (226) -- red giant stars (1372) -- stellar evolution (1599) -- stellar evolutionary models (2046) -- asymptotic giant branch stars (2100)}

\section{Introduction} \label{sec:intro}

Lithium--rich red giants have been a long--standing mystery in stellar astrophysics. From our current understanding of stellar evolution, surface lithium is depleted in a star when the star goes through the first dredge--up (FDU) phase before becoming a red giant. However, \app1\% of giants are found to be Li--rich, $\rm A(Li)\geq 1.5$ dex \citep[e.g.,][]{gao_2019, casey_2019, Martell2021, sayeed_2024}. Since their discovery by \cite{WallersteinSneden1982}, several theories have been proposed to explain their existence, such as enhancement due to \mt from asymptotic giant branch (AGB) stars \citep[e.g.,][]{SackmannBoothroyd1992}, tidal spin--up from a companion \citep[e.g.,][]{casey_2019}, engulfment of planets or sub--stellar companions \citep[e.g.,][]{Alexander1967, SiessLivio1999a, SiessLivio1999b, VillaverLivio2009, adamow_2012, King1997, Israelian2004, Israelian2009, DelgadoMena2014, Martell2021}, mixing induced by the core He--flash \citep[e.g.,][]{kumar_2020, Zhang2021, Mallick_2023}, thermohaline mixing \citep[e.g.,][]{Lattanzio2015}, and others. While large--scale spectroscopic surveys such as \lamost \citep[][]{lamost_2012, cui2012} and \galah \citep[][]{galah_survey, hermes_galah, galah_sven_2021} have enabled comprehensive, population studies into the formation of these objects  \citep[e.g.,][]{gao_2019, wheeler2021, singh_2019, yan_2021, Ming-hao_2021, yan_2022, zhou_2022, deepak_reddy_2019, Gao2020, kumar_2020, deepak_2020, Martell2021, melinda_2021, chaname_2022}, their formation mechanism is still uncertain. 

Lithium can be observed at the surface of stars when the Cameron--Fowler (CF) mechanism \citep{cameron_1971} is active. This is when the following proton--proton reaction,
\begin{equation}
    \rm{^3He} + \rm{^4He} \rightarrow \rm{^7Be}
\end{equation}
produces beryllium ($^7$Be), which is then transported to cooler regions to create $^7$Li via electron capture. Depending on the timescale and nature of the mixing, the \li may then be transported to a hydrogen burning region and destroyed, via proton capture. Lithium is also theorized to form via Hot Bottom Burning (HBB) in the envelopes of intermediate--mass ($4-8$ \solmass) AGB stars \citep{SackmannBoothroyd1992}. HBB, an example of CF mechanism, occurs when the bottom of the convective layer reaches a temperature of 40 $\times10^6$ K, initiating proton--capture nucleosynthesis at the base of the outer envelope \citep[][]{Garcia2013}. 

Red giants can therefore achieve \li enhancement through intrinsic (internal production and transport to the surface) or extrinsic (\mt or engulfment) means. Tidal spin--up from a binary companion is a means of initiating the CF mechanism inside a star, producing \li that is transported to the stellar surface \citep[][]{casey_2019}. Similarly, the onset of core He--flash may cause mixing and initiate the CF mechanism as well \citep[e.g.,][]{kumar_2020, Zhang2021, Mallick_2023}. Alternatively, both planet engulfment and \mt from a Li--producing intermediate--mass AGB companion would enrich the red giant in lithium. Regardless, both intrinsic (with the exception of the He--flash scenario) and extrinsic (ie. tidal spin--up or \mt from an AGB) channels require the presence of a companion.



Recently, \cite{sayeed_2024} analyzed a large sample of \lirich giants in \galah and found no obvious indications of binarity. Their analysis utilized \gaia radial velocity errors and RUWE (re--normalized unit weight error) measurements to investigate binarity of \lirich giants \citep[][]{gaia_collab, gaia_2018, Lindegren_2018, chance_2022}. Given the similarities in the distribution of \gaia RUWE values and radial velocity errors between samples of \lirich and \lin samples, they concluded that \lirich giants are not preferentially in binary systems with underlying architectures that would lead to signals in these metrics. However, \gaia RUWE is sensitive to binaries at wide separations ($\gtrsim 2 \rm \; AU$) assuming circular orbit and all else being equal, while excess noise in radial velocity errors arises from binaries with close separations ($\lesssim 1 \rm \; AU$). Because the astrometric selection suggests that the \lirich giants are not in wide binaries, we investigate the possibility that they are in closer binaries where astrometry is less sensitive and there is not an obvious companion in the photometric data; a white dwarf--red giant binary is a natural choice since this satisfies both criteria. In order to test for evidence of binarity in radial velocity data, or rather to rule out binarity or subsets of architectures, it is imperative to understand the parameter space that these objects would occupy.

In addition to astrometry, elemental abundances can also be used to probe binarity. For instance, chemically peculiar stars enhanced in \sprocess elements (or slow neutron--capture process) have previously been proposed to at one time host an AGB companion. Intermediate--mass stars can produce \sprocess elements in their interiors which are transported to the stellar surface via mixing episodes \citep[e.g.,][]{Wood1983, Herwig2005, Garcia2006, Garcia2009, Karakas2012, Karakas2014_review, Cseh_2018, Norfolk_2019, Cseh_2022, denHartogh_2022, Escorza_2023}. \cite{sayeed_2024} found \lirich stars at the base of the red giant branch are slightly enhanced in \sprocess elements compared to a reference set of Li--normal stars at the same evolutionary state, such as barium, $\Delta(\rm{Ba}) = 0.13 \pm 0.05$ dex. They suggest that this enhancement in both \sprocess and Li could be caused by enrichment via \mt from an intermediate--mass AGB. Binary interactions could also show signatures in stellar activity, although no direct signal of this has been found thus far \citep[e.g.,][]{julio2024, sayeed_2024, rolo2024}.  

\begin{figure}[t!]
    \centering
    \includegraphics[width=\linewidth]{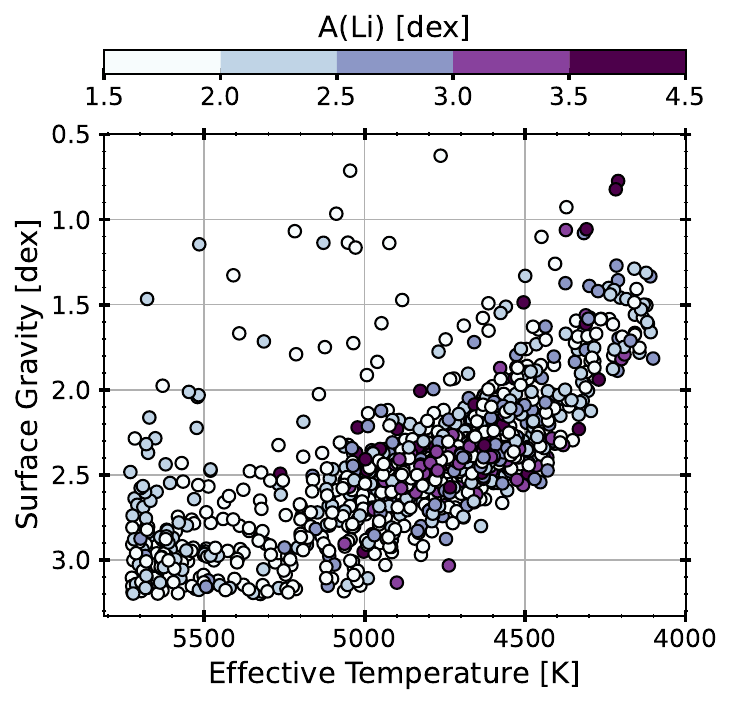}
    \caption{Kiel diagram of \lirich giants in \galah coloured by their Li abundance with selection criteria from \cite{sayeed_2024}.}
    \label{fig:hrd}
\end{figure}

In an effort to better understand the complex and likely diverse contributions to Li--enhancement, our study aims to put constraints on one of the hypothesized pathways: assuming binarity as a plausible origin of Li--enrichment in red giants, we explore the parameter space of possible underlying binary systems using theoretical models. We first perform an in--depth exploration of binarity in Section \ref{sec:cosmic_results}, using the binary population synthesis code \cosmic \citep{Breivik2020}, with initial conditions constrained by \galah observations. We then examine the parameters of systems that could lead to red giants with Li--enrichment. Second, we use stellar models and nucleosynthetic yields, and compare to observed abundance measurements to determine possible AGB progenitor properties in Section \ref{sec:models_results}. We then connect the binary synthesis and nucleosynthesis in pinpointing likely architectures that may lead to the observed fraction of \lirich red giants captured in Section \ref{sec:discuss}.  

\section{Methods} \label{sec:methods}

\subsection{\galah Sample}
We use the \lirich sample from \cite{sayeed_2024} -- hereby referred to as S24 -- to initialize the properties of red giants to model with the binary population synthesis software, \cosmic (see Section \ref{sec:cosmic} for description). We provide a brief description of the sample selection below, but refer the reader to the original paper for more details. S24 used the latest data release (DR3) of the Galactic Archaeology with HERMES survey \citep[\galah,][]{galah_survey, hermes_galah} to identify \lirich red giants; we use the same data set to maintain consistency between our analysis. \galah is a spectroscopic survey that provides high--resolution ($R \approx 28,000$) spectra in four wavelength bands ($4713-4903$, $5648-5873$, $6478-6737$, and $7585-7887$ \ang). \galah DR3 \citep{galah_sven_2021} includes one--dimensional spectra, stellar atmospheric parameters, and up to 30 individual element abundances for 588,571 nearby stars. S24 identified 1455 \lirich giants in \galah with $\rm A(Li) \geq 1.5$ dex\footnote{\ali = [Li/Fe] $+$ [Fe/H] $+$ 1.05}, \teff $=[3000, 5730]$, \logg $=[-1, 3.2]$ dex, and good quality flags in the spectrum, [Fe/H] measurements, and more (see Section 2.2 in S24). They also performed a \dg analysis where they identified a Li--normal (or \dg) star for each \lirich star where a \dg is at the same evolutionary state as the \lirich star (based on \teff, \logg, [Fe/H], [Mg/Fe]), but with $\rm A(Li)< 1.0$ dex. 
For this work, we use the full sample of 1455 \lirich stars to initialize red giant properties in \cosmic. The sample is shown in Figure \ref{fig:hrd} on a Kiel diagram, coloured by absolute lithium abundance, \ali, from \galah.

\subsection{\cosmic} \label{sec:cosmic}
\cosmic (Compact Object Synthesis and Monte Carlo Investigation Code) is a binary population synthesis suite that simulates binary populations from zero--age main sequence, through any potential binary interactions, up to the formation of double stellar remnants \citep{Breivik2020}. While the primary use of \cosmic has been to forecast and interpret double stellar remnant populations and their mergers \citep[e.g.,][]{Zevin21, Wong21, Thiele23, Wong23}, \cosmic can also be used to simulate earlier phases of evolution where only one stellar remnant has formed \citep[e.g.,][]{Breivik17,Chawla22,El-Badry23}, or both stars are still fusing hydrogen \citep[e.g.,][]{Leiner21}. Here, we consider the binary progenitors that can produce \lirich giants through stable Roche--Lobe overflow (RLOF). We do not consider stellar wind accretion due to its inefficiency; we discuss the limitations of this method of enrichment in Section \ref{sec:limits}.  

\begin{figure}
    \centering
    \includegraphics[width=\linewidth]{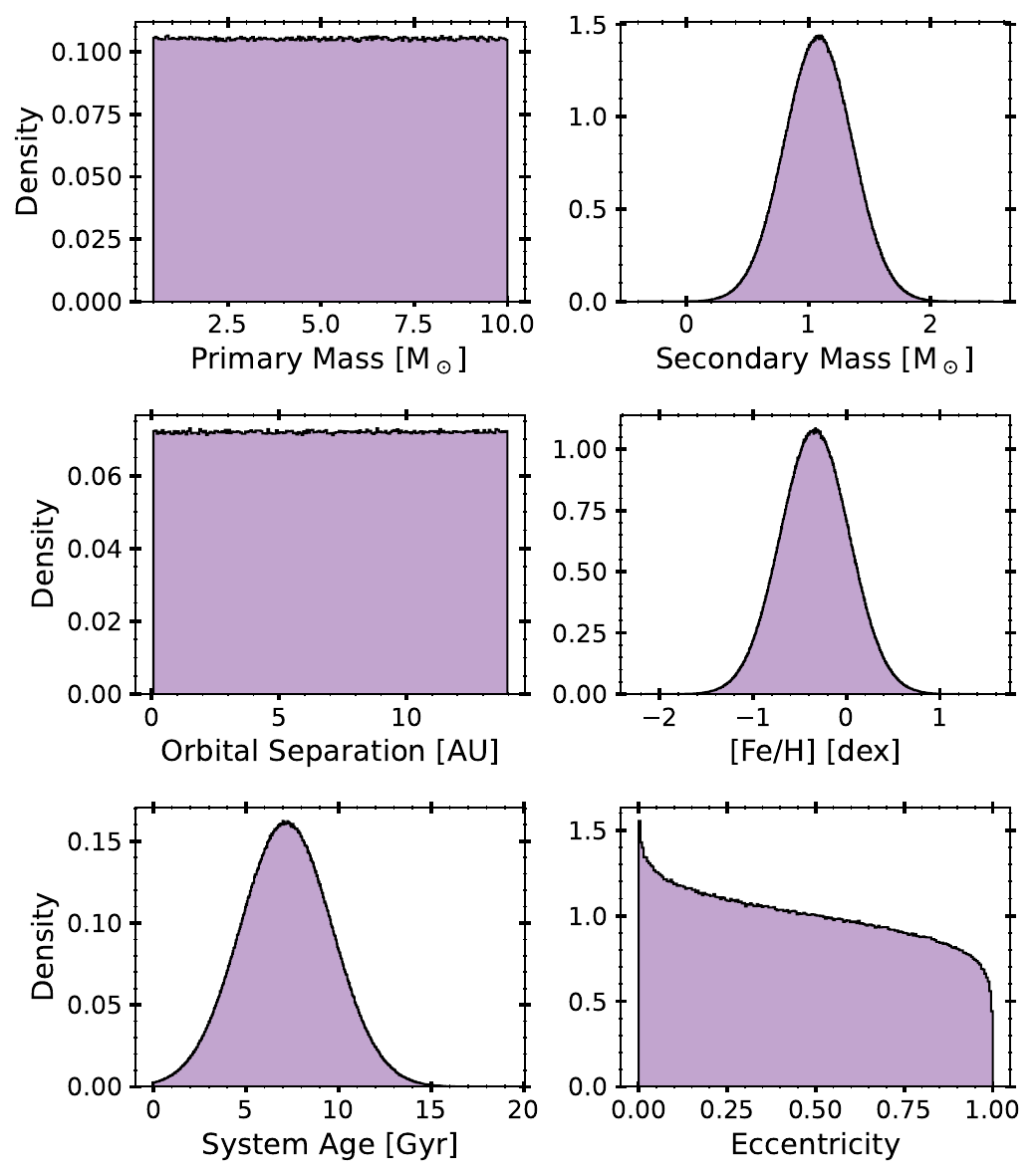}
    \caption{Input system parameters used to initialize 9 million binary systems in \cosmic. The primary mass and orbital periods were sampled from a uniform distribution, while the secondary mass, metallicity, and system age were sampled from a Gaussian distribution from \galah observations.}
    \label{fig:input}
\end{figure}

We initialize 8,995,738 million binary systems with \cosmic, generated following \galah's distribution of stellar parameters. Systems are randomly sampled from a normal distribution for age in Gyr ($\mu=7.2, \; \sigma=2.5$) and metallicity in dex ($\mu=-0.34, \; \sigma=0.37$) following \galah observations of 112,620 red giants. The initial separation in AU is sampled from a uniform distribution (in log space) where the minimum separation is twice the average Roche--Lobe radius of the white dwarf \citep{Eggleton1983}, and the upper bound was set to 3000 \solrad or 14 AU, chosen arbitrarily but above 10 AU. The secondary's mass (ie. red giant) is sampled from a normal distribution of ZAMS (zero--age main--sequence) mass in \solmass ($\mu=1.1,\; \sigma=0.3$); however, the primary's mass (ie. AGB companion now a white dwarf) is sampled from a uniform distribution between $0.5-10$ \solmass. We discuss the implications of our choice of prior for the primary's mass and the metallicity in Section \ref{sec:assumptions}. We evolve each system using \cosmic until the system reaches a final age randomly sampled from a normal distribution generated from observed red giant ages in \galah. Figure \ref{fig:input} shows the distribution of all input parameters used to initialize \cosmic systems.

\begin{figure*}
    \centering
    \includegraphics[width=\linewidth]{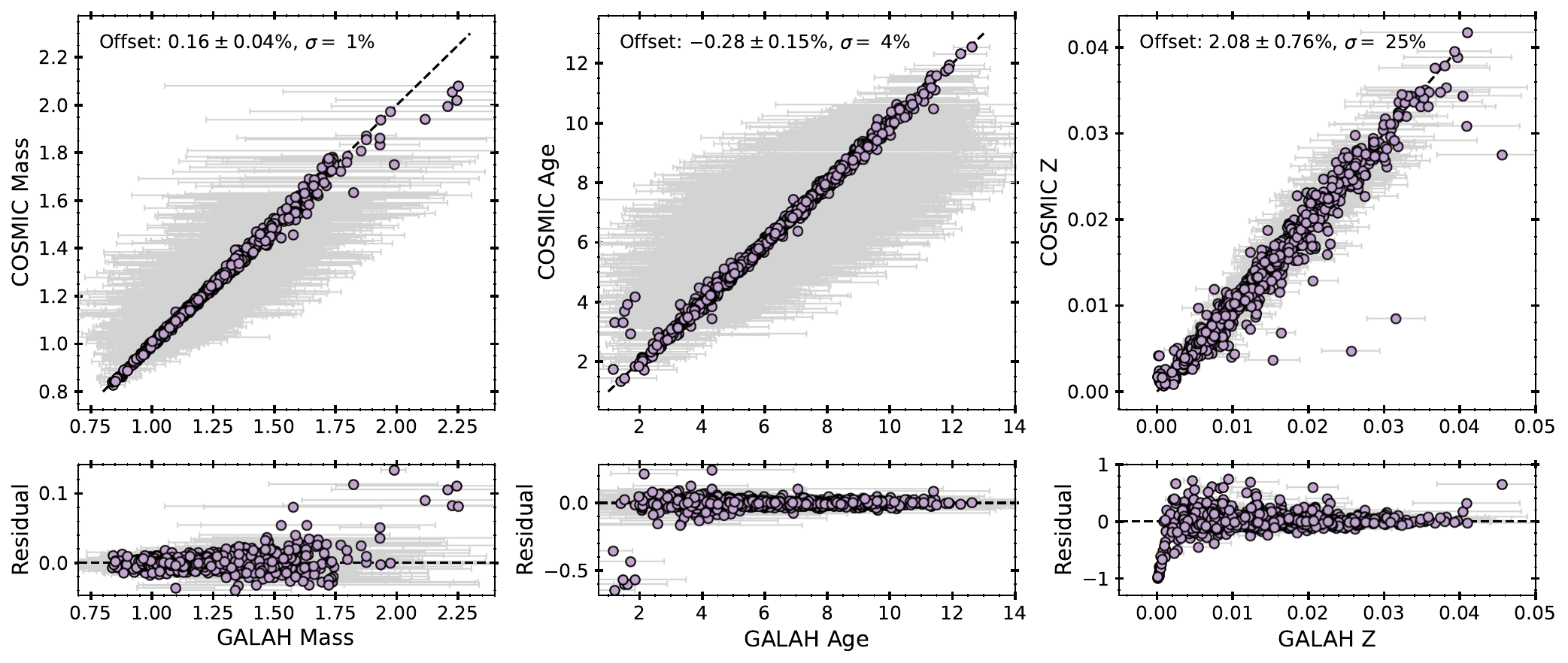}
    \caption{Comparing \galah red giant parameters to \cosmic system parameters after matching on mass, age, and absolute metallicity, $Z$. The text indicates the mean offset with error, and the scatter. The bottom panels show the fractional difference.}
    \label{fig:match}
\end{figure*}

We evolve the population with single and binary evolution models following the \cosmic\ default assumptions\footnote{https://cosmic-popsynth.github.io/COSMIC/inifile/index.html} with solar metallicity set to $Z_{\odot}=0.019$. The model assumptions that are particularly important for constraining the impact of binary interactions in the \lirich giants are those for wind mass loss and accretion, the stability of \mt, the accretion efficiency during stable \mt, and the common envelope (CE) ejection efficiency. For wind mass loss, we apply the standard prescriptions applied in \cite{Hurley2002}; however, the mass loss prescription for the AGB is coded from \cite{VassiliadisWood1993} which considers the rapid mass loss in AGB stars before the super--wind phase. We add metallicity--dependent prescriptions for massive stars \citep{Vink2001} and we fix wind velocities to the prescription of \cite{Belczynski2008}. The stability of \mt is assumed to follow \cite{Hurley2002} where the critical mass ratio of the donor to the accretor mass is three for donors with radiative envelopes, and follows \cite{Hjellming1987} for stars with convective envelopes. In the case of stable \mt, we assume that the accretor can accept mass at ten times its thermal rate. Finally, we assume a CE ejection efficiency of unity such that $100\%$ of the initial orbital energy can be used to unbind the donor star's envelope.

From the 9 million evolved systems, we selected systems consisting of a white dwarf and a red giant at the last time--step of the \cosmic evolution. This is to ensure the age distribution of simulated red giants is consistent with the age distribution of observed red giants in \galah. We also isolated systems such that the white dwarf star must undergo an AGB phase during its evolution; this means we only consider carbon--oxygen and oxygen--neon white dwarfs. For 9497 systems, we end up with a final system consisting of a white--dwarf and red giant. The overall rate of binaries from \cosmic is therefore $0.001$ white--dwarf -- red giant systems per solar mass.

We characterize each candidate system by its red giant mass, system age (ie. age of the red giant), and metallicity, and also characterize the \lirich giants in S24 by their mass, age, and metallicity provided by \galah. Then, we select evolved \cosmic candidate systems with a red giant that is the closest match to \galah observed \lirich red giants. We do this by minimizing the $\chi^2$ distance in the parameter space of mass, age, and metallicity:
\begin{equation}
    \min \sum_{i=1}^4 \frac{(X_{\rm GALAH,i} - X_{\rm COSMIC, i})^2}{\sigma_{\rm GALAH,i}^2}
   \label{chi2metric}
\end{equation}
where $X_{\rm GALAH,i}$ and $X_{\rm COSMIC,i}$ is the value for the $i$th stellar parameter for the \lirich star in \galah or in the candidate \cosmic binaries respectively, and $\sigma_{\rm GALAH,i}$ is the uncertainty in the parameters provided by \galah. The \galah \lirich red giants and the red giant candidates in \cosmic already share a similar parameter space before the matching. Hence, repeated match cases exist where one \cosmic red giant is equidistant to two or more \galah \lirich red giants in our parameter space. For each case of repeated match, we retain the match for the first \galah star. Then, we rematch the other \galah star(s) with a pool of \cosmic candidates that did not get a match in the first round. To ensure high fidelity, we require that the difference in each parameter used for matching is less than the mean error in \galah for that parameters. The mean error in \galah for mass, age, and [Fe/H] is $0.2 \rm \; M_\odot$, 2.5 Gyr, and 0.02 dex, respectively. Ultimately, we recover 1104 unique \cosmic systems that match well with \galah \lirich red giants in our parameter space. Figure \ref{fig:match} shows the stellar parameters -- mass, age, and absolute metallicity -- for matched systems, \galah \lirich red giants to \cosmic systems.

\subsection{Monash Stellar Models}

We use detailed low and intermediate--mass stellar evolution models from \citet{karakas2014helium} and \citet{karakas2016stellar} with initial masses between $M_{\rm ZAMS} = 4 - 8 \rm \; M_\odot$. The structural evolution of the models from \citet{karakas2014helium} was calculated with the Monash stellar evolution code (an adaptation of the Mt. Stromlo code, \citealt{LattanzioThesis, Lattanzio86, Frost96, Karakas07}), using a limited nuclear network with isotopes that contribute significantly to energy generation ($^{1}$H, $^{3}$He, $^{4}$He, $^{12}$C, $^{14}$N and $^{16}$O). The evolution of the models extend from the pre--main sequence to the tip of the thermally pulsing AGB (TP--AGB), with an initial composition approximately solar ($X_{\rm i} =  0.706, Y_{\rm i} = 0.28, Z_{\rm i} = 0.014$; \citealt{Asplund_2009}). The input physics required for these calculations are discussed in detail in \citet{Karakas2014a}; briefly, low temperature opacity tables are from \citet{marigo2009low} and high temperatures from OPAL \citep{Iglesias96}. Convection is treated using the Mixing Length Theory \citep{Vitense53, Prandtl25} with an efficiency parameter of $\alpha_{\rm MLT} =1.86$, and borders between convective and radiative regions are solved using the algorithm described in \citet{Lattanzio86}. Rates of mass loss on the AGB are calculated using the prescription from \citet{vassiliadis1993evolution}, and no mass loss is included on the red giant branch. In \citet{karakas2016stellar}, the evolution models are post--processed (see \citealt{Cannon93, Lattanzio96, Lugaro12}) to extend the nuclear network from 6 to 328 species and follow \sprocess reactions. The reaction rates used for the larger network were adopted from the JINA reaclib database\footnote{\url{https://reaclib.jinaweb.org/}} and updates from \citet{karakas2016stellar}. The $\rm ^{13}C(\alpha,n)^{16}O$ reaction provides the dominant source of free neutrons in low mass stars; pockets of $^{13}$C were artificially created in the calculations by inserting protons into the top of the He shell (see Section 2.2 of \citealt{karakas2016stellar}). This creates partially mixed zones where protons can be captured by $^{12}$C nuclei to form $^{13}$C (with mass extent $10^{-3} \rm \; M_\odot$ for the 4.25 and $4.5 \rm \; M_\odot$ models, and $10^{-4} \rm \; M_\odot$ for 4.75 and $5 \rm \; M_\odot$). No pockets are included for the $5.5-8$ \solmass models as temperatures are too high for a reasonable source of free protons to exist (see \citealt{Goriely04, herwig2004dredge}).

The chemical makeup of the mass expelled from low and intermediate--mass models to the interstellar medium over their lifetime (or their ‘yield’) is governed by mass and composition dependent nucleosynthesis and mixing processes. Some of the richest of these processes, i.e., those that give rise to considerable changes in surface abundance, occur on the TP--AGB. The TP--AGB comprises only a few percent of the total evolutionary lifetime for these models ($0.3-0.4\%$)\footnote{From Tables 1 and 2 in \citet{karakas2014helium}.} and is characterized by recurrent thermonuclear runaway events (or thermal pulses) that originate from the He--burning shell. The energy released from thermal pulses can drive the outer regions of the model to expand and allow the convective envelope to move inwards in mass, into regions of prior He--burning and \sprocess nucleosynthesis. These large scale convective mixing episodes are known as the third dredge--up (hereafter, TDU) and can occur after each pulse. As a result, the stellar surface may be polluted tens to hundreds of times (depending on initial composition and $M_{\rm ZAMS}$) in a short evolutionary timeframe. For example, the $4.5 \rm \; M_\odot$ model experiences TDU episodes after 30 of its 31 total thermal pulses, during the last $\rm 0.5 \; Myrs$ of the total $\rm 135 \; Myr$ lifetime. 

\begin{figure*}[t!]
    \centering
    \includegraphics[width=0.32\textwidth]{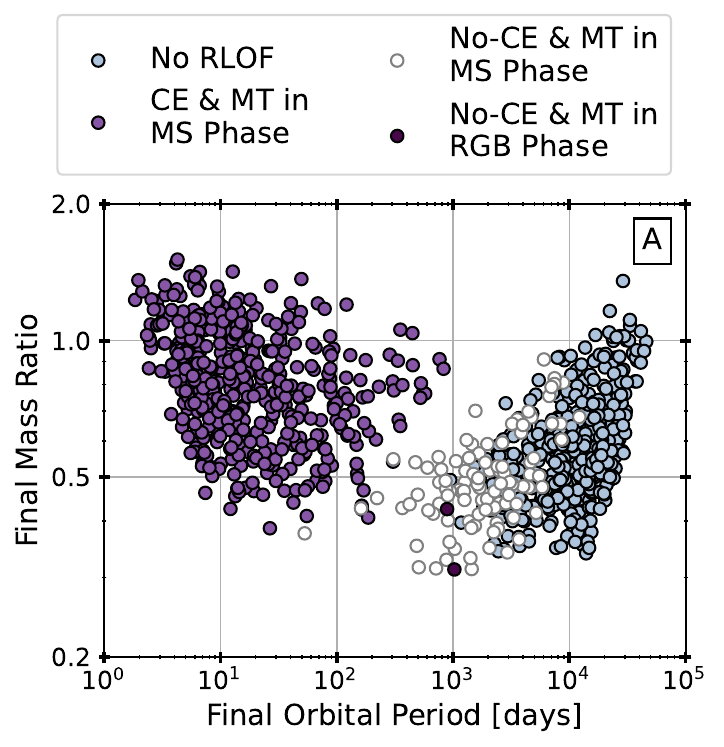}
    \includegraphics[width=0.32\textwidth]{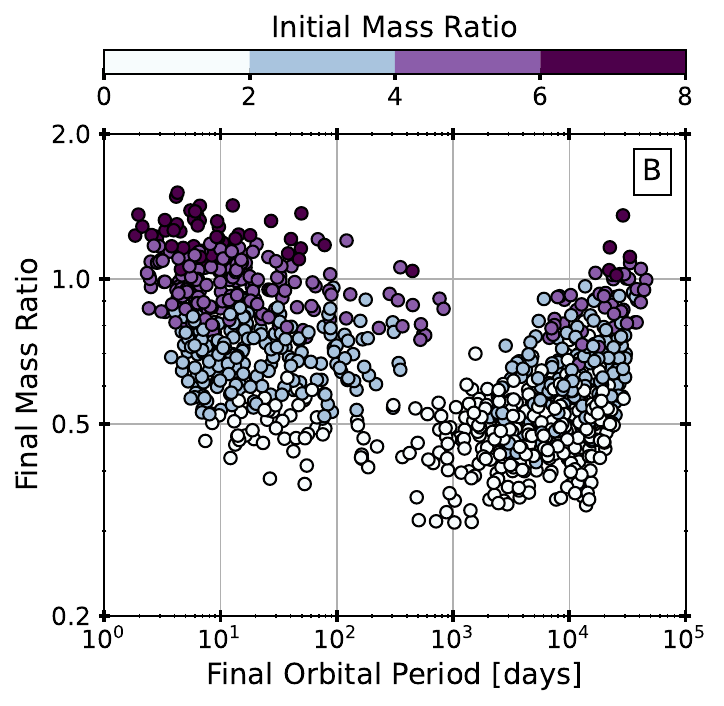}
    \includegraphics[width=0.32\textwidth]{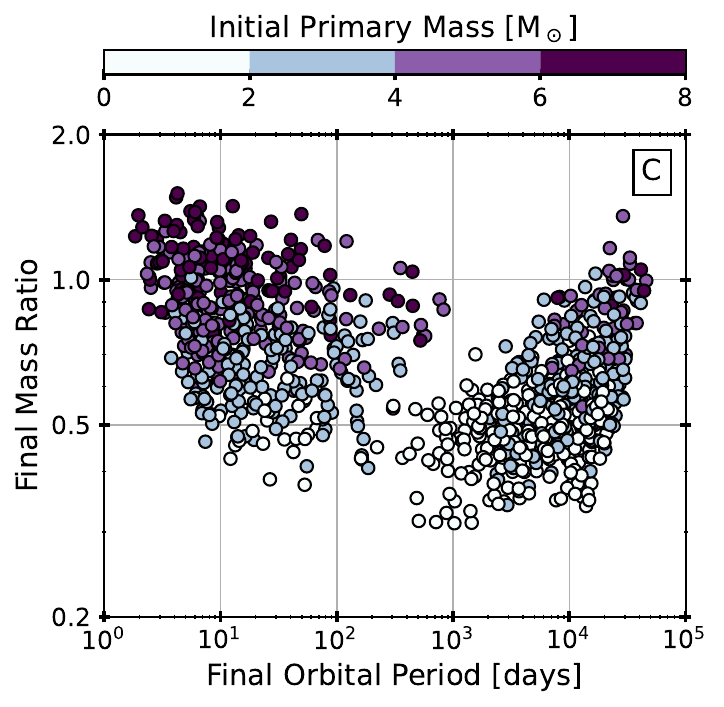}
    \caption{\textit{Left:} Final mass ratio and orbital period (in days) of the \cosmic systems, where mass ratio is defined as $q=M_1/M_2$ where $M_1>M_2$ ($M_1$ is now the white dwarf and $M_2$ is the red giant). The blue points indicate systems that do not undergo RLOF. The purple points indicate systems with a CE phase and \mt (MT) when the secondary is on the main--sequence (MS). The white and black points indicate systems without a CE phase where \mt occurs when the secondary is on the MS or on the RGB, respectively. 
    \textit{Middle:} Final mass ratio and orbital period of the \cosmic systems coloured by the initial mass ratio. 
    \textit{Right:} Same as earlier panels, but coloured by initial mass of the primary, $M_1$. }
    \label{fig:qi}
\end{figure*}

The dominant free neutron source in intermediate--mass stars derives from the $\rm ^{22}$Ne$(\alpha,n)^{25}$Mg reaction, which has a lower total integrated neutron flux than the dominant primary $^{13}$C source in low mass models (see e.g. \citealt{lugaro2023s,arcones2023origin}). A weaker neutron flux favours the formation of the lighter first magic peak nuclei, Sr, Y, and Zr, over the second and third magic peaks, Ba, La, Ce, Pr, Nd, and Pb. Further, the surfaces of intermediate--mass AGB stars are more difficult to pollute with TDU than their lower mass counterparts given that they have larger convective envelopes and smaller intershell regions. For example, the shift from $4.5  \rm \; M_\odot$ to $5 \rm \; M_\odot$ at solar metallicity results in a convective envelope \app$13$\% more massive at the 10th thermal pulse. The same shift in ZAMS mass results in a decrease of the maximum mass of the flash--driven convective region (see Figure 14 from \citealt{Karakas2014_review}) by \app$14$\% in the $5 \rm \; M_\odot$ model, with 22\% less material mixed into the envelope following TDU. Any fresh $^{12}$C mixed to the surface can be destroyed during quiescent periods of H shell--burning between pulse cycles, where temperatures in intermediate--mass stars can be sufficient ($T \gtrsim 50 \rm \; MK$) to promote H--burning reactions at the base of the convective envelope during HBB \citep{lattanzio1992hot,boothroyd1995hot,mcsaveney2007abundances}, and can increase surface Li through the second chain of the proton--proton cycle (\textit{pp--}cycle; \citealt{bethe1938formation}), 
\begin{equation}
   \rm p(p,e^+\nu)d(p,\gamma)^{3}He(\alpha,\gamma)^{7}Be(e^-,\nu)^{7}Li. 
\end{equation}
The convective region in which HBB occurs allows the $^{7}$Be to be mixed up into cooler regions to undergo electron capture before it can be destroyed by proton capture \citep{smith1989synthesis,smith1990occurrence}. The magnitude of $^{7}$Li production during HBB is dependent on the availability of $^{3}$He fuel; no initial abundance of $^{3}$He is assumed at the ZAMS in the \citet{karakas2014helium} models, where any $^{3}$He present in the envelope at the first thermal pulse is the product of proton--proton reactions on the main sequence and mixed to the surface during the FDU. 

\section{Results}\label{sec:results}
We describe the results from \cosmic simulations in Section \ref{sec:cosmic_results}, and results from stellar models in Section \ref{sec:models_results}. We compare the results from \cosmic and stellar models in Section \ref{sec:cosmic_models}, and from \galah observations and stellar models in Section \ref{sec:galah_models}.

\subsection{Results from Binary Star Models} \label{sec:cosmic_results}

Final systems in \cosmic satisfy the following conditions: 
\begin{enumerate}[i)]
    \item the system consists of a red giant and white dwarf\footnote{\texttt{kstar\_1==11.0} or \texttt{kstar\_1==12.0} and \texttt{kstar\_2==3.0} or \texttt{kstar\_2==4.0}; Column names and descriptions can be found here: \url{https://cosmic-popsynth.github.io/COSMIC/output_info/index.html}}
    \item the primary must have undergone an AGB phase\footnote{\texttt{kstar\_1==5.0} or \texttt{kstar\_1==6.0}}
\end{enumerate}
Of the 1104 unique \cosmic systems matched to a \lirich giant in \galah, 92\% (1021/1104) of the systems host an AGB star that went through thermal pulses. Of these, 55\% or 559 systems underwent RLOF\footnote{\texttt{evol\_type==3.0}}, where 82\% (457/559) experienced a CE phase,\footnote{\texttt{evol\_type==7.0}} and 18\% (102/559) did not. This left 45\% or 462 systems with no RLOF that remained in binary systems with no coalescence. 

Figure \ref{fig:qi}A shows the final mass ratio and orbital period from the \cosmic simulation, where we differentiate between systems that underwent RLOF with a CE, without a CE, and those that did not undergo RLOF. Mass ratio is defined as $q=M_1/M_2$ where $M_1>M_2$ initially. The primary ($M_1$) is now the white dwarf that has gone through the AGB phase, and the secondary ($M_2$) is now observed as a red giant. Figure \ref{fig:qi}A also indicates if \mt occurred during the main--sequence (MS) phase or the red giant phase of the accreter (ie. red giant). Of the 559 systems that underwent RLOF, \mt occurred during the red giant phase of the accretor for only two systems, and during the main--sequence phase for the remaining 557 systems. Finally, \app6\% of the systems that underwent stable RLOF experienced a mass ratio reversal where the primary mass (the initially more massive component) fell below the secondary mass before \mt, widening the orbit in 36 systems and narrowing the orbit in one system. Furthermore, while the input eccentricity of our systems are distributed randomly, all systems that undergo RLOF have a final eccentricity of 0. This is because all binaries which undergo RLOF \mt in \cosmic are assumed to circularize through strong tides before the Roche--Lobe is filled. 


Figure \ref{fig:qi}B shows the final system properties coloured by the initial mass ratio for our systems. We notice a gradient where systems with initial higher mass ratio tend to have near equal final mass ratios. Figure \ref{fig:qi}C shows the same systems coloured by the initial mass of the primary (ie. AGB). We find 29\% of our systems (296/1021) qualify for HBB to take place given that only these systems consist of an intermediate--mass AGB (ie. $M = 4.25-8$ \solmass). These systems have a mean separation of $3.3 \pm 0.5 \rm \; AU$, and initial mass ratios between $q_i = 0.5-1.6$.


Based on Figure \ref{fig:qi}, our \cosmic results can be summarized as the following:

\begin{enumerate}[i)]
    \item Roche--Lobe Overflow with CE phase: end up in short period systems ($0.04-2.25, \; \mu=0.25$ AU) and initially have a range of initial mass ratios ($q_i=1-8$). In 93\% of the systems, the accretor gains mass.
    
    \item Roche--Lobe Overflow with no CE phase: end up in longer period systems ($0.3-14.0, \; \mu = 4.4 $ AU), and most systems (69\%) have initial mass ratios $q_i < 2$. However, this mass ratio is an input into \cosmic such that mass ratios $q \gtrsim 1-2$ go into \ce. 
    
    \item No Roche--Lobe Overflow: end up in longer period systems ($2-33, \; \mu = 13$ AU), and have a range of initial mass ratios, $q_i = 0.4-8$. 
\end{enumerate}

\begin{figure}[t!]
    \centering
    \includegraphics[width=\linewidth]{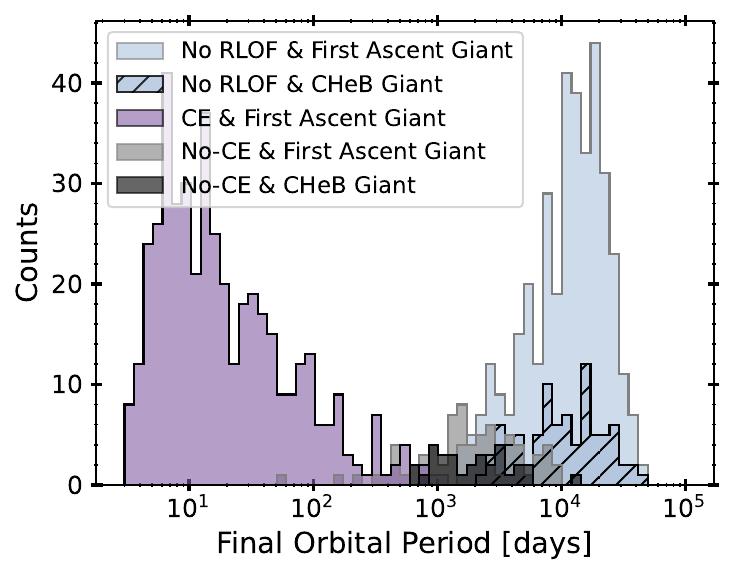}
    \caption{Final separation of systems where we distinguish between systems with a first ascent red giant and those with a core helium--burning red giant.}
    \label{fig:rc_vs_rgb}
\end{figure}

\cosmic also differentiates between the first ascent and \cheb phase of the red giant. Of the 559 systems that undergo RLOF, the secondary was a \cheb giant for only 6\% of the systems (33/559). Interestingly, for systems with a \ce, the secondary is a first ascent giant for 99\% of systems. In Figure \ref{fig:rc_vs_rgb}, we show the final orbital period of the systems separated into whether the system produced a CE and whether the red giant was a first ascent giant or a \cheb giant; we find no difference in the final separation between systems with a first ascent or \cheb giant. 

\cosmic provides stellar properties before, during, and after evolution of each system. We can therefore use the mass accreted by the red giant as well as its envelope mass to predict the amount of lithium transferred to the red giant for a given initial Li abundance. Using Equation \ref{eq:formula}, we can derive the giant's abundance post mass transfer given the mass fraction of the red giant after accretion ($X_\textrm{RGB,a.a.}$),

\begin{equation}
    X_\textrm{RGB,a.a.} = \frac{ X_\textrm{RGB, b.a.} \cdot M_\textrm{env}+X_\textrm{AGB}\cdot M_\textrm{acc}}{ M_\textrm{env}+M_\textrm{acc} }
    \label{eq:formula}
\end{equation}
where $X_\textrm{RGB, b.a.}$ is the surface mass fraction of the red giant before accretion (b.a.),  $X_{\rm AGB}$ is the surface mass fraction of AGB donor, $M_\textrm{env}$ is the RGB envelope mass, and $M_\textrm{acc}$ is the accreted mass. In Section \ref{sec:calculation}, we outline our methods for calculating the surface abundance of the red giant star post \mt.

We use \cosmic results, \galah observations, and stellar theory to place upper limits on each of the components in Equation \ref{eq:formula}. \cosmic results show that for systems that experienced RLOF with a TP--AGB, the mass accreted by the RGB is $< 0.5$ \solmass, while the RGB envelope mass immediately before the RLOF phase begins is $< 1.5$ \solmass. Therefore in Equation \ref{eq:formula}, $M_\textrm{env} = 0-1.5$, and $M_\textrm{acc} = 0-0.5$. We use Equation \ref{eq:formula} to simulate expected Li abundance of an RGB ($X_\textrm{RGB, a.a.}$) with a range of envelope masses and accreted masses after \mt from an AGB. 

\begin{figure}[t!]
\centering
\includegraphics[width=\linewidth]{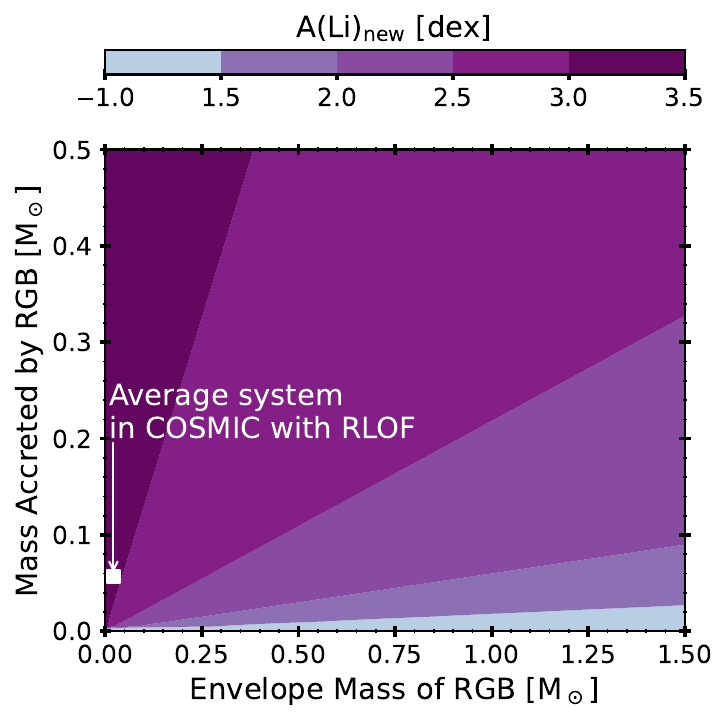}
\caption{The expected \ali of a red giant after \mt from an AGB for a range of accreted masses and RGB envelope masses in \solmass. The red giant has $A\rm(Li)_{RGB} = -0.4$ dex before accretion and the AGB has $A\rm(Li)_{AGB} = 3.2$ dex. The colour denotes the new Li abundance, \ali. The square marker shows the mean envelope mass of red giant, and mean mass accreted by the red giant in \cosmic.}
\label{fig:calcs}
\end{figure}

Figure \ref{fig:calcs} shows the final \ali of a red giant after accreting mass from an AGB companion where $A\rm(Li)_{RGB} = -0.4$ dex before accretion and $A\rm(Li)_{AGB} = 3.2$ dex. The white square marker shows the average RGB envelope mass and accreted mass from \cosmic simulation. Figure \ref{fig:calcs} suggests that mass accreted by an RGB has a more significant effect than the RGB's envelope mass on the resulting Li--abundance. For instance, an RGB can achieve $\rm A(Li)> 3.0$ dex even with an envelope mass of \solmass $< 0.05$. Figure \ref{fig:calcs} also confirms that our observed \ali values can be re--produced given \cosmic results assuming realistic initial Li abundances.

\subsubsection{Comparison with Binaries in \apogee}

By comparing our results in \cosmic to existing multi--epoch data, we can learn which of our binary architectures in \cosmic would not be detected, and which in principle can lead to \lirich giants. We compare our results from \cosmic to known binaries in \apogee Data Release 16 (DR16). \apogee, a component of Sloan Digital Sky Survey IV \citep[SDSS IV,][]{Gunn2006, Blanton2017}, provides multi--epoch, high resolution ($R\sim22,500$) infrared ($H-$band) spectroscopy for targets in the Northern and Southern hemispheres. \cite{apw_2020} used a custom Monte Carlo sampler called \textit{The Joker} to discover and characterize binary systems with radial velocities in \apogee DR16. They performed quality cuts, such as removing sources with less than three visits, to produce a catalog of \app20,000 systems. The \apogee binary catalog from \cite{apw_2020} contains 2829 sources also observed by \galah\footnote{after restricting the sample based on the criteria in \cite{sayeed_2024}}, where 1176 are red giants\footnote{\teff $=[3000,5730]$ K, \logg $=[-1,3.2]$ dex}, and 13 are \lirich giants. These \lirich giants in \apogee all have three visit spectra ($N_{\rm visit} = 3$) and a baseline of 37.9 days. In comparison, 3104 targets were observed by both \apogee DR16 and \galah DR3 with at least three visits, of which 1274 are red giants, and 15 are \lirich giants. The fraction of red giants in \galah and \apogee (with $N_\textrm{visit} \geq 3$) is $\sim 41\%$ (1274/3104) and the fraction of \lirich giants is $\sim 1.1\%$ (15/1274). These fractions are similar to sample overlap between \galah and the \apogee binary catalog, where $\sim 41\%$ (1176/2829) are red giants and $\sim 1.1\%$ (13/1176) are \lirich giants. Therefore, \lirich giants in \apogee are not preferentially binaries. \cite{apw_2020} also provide a \textit{Gold Sample} of 1032 systems that have converged, unimodal, or bimodal posterior samplings -- unlike most of the catalog -- and pass a strict set of quality cuts (such as $N_\textrm{visit} > 5$). While there are 17 \galah stars in the \textit{Gold Sample}, none are \lirich giants.  

\begin{figure}[t!]
    \centering
    \includegraphics[width=\linewidth]{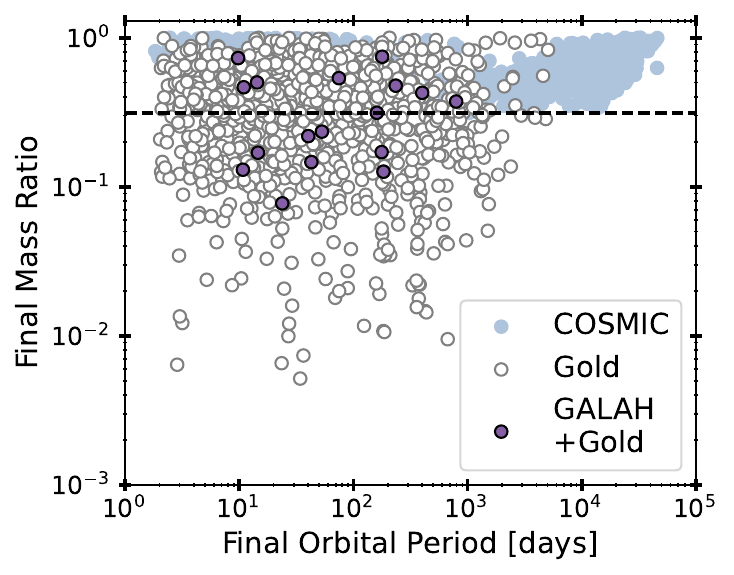}
    \caption{The binary parameters from \cosmic (blue), the \textit{Gold Sample} in \cite{apw_2020} (white), and overlapping \galah stars in the \textit{Gold Sample} (purple). The black dashed line shows the lower limit of our \cosmic systems.}
    \label{fig:apogee_cosmic_ruwe}
\end{figure}

Figure \ref{fig:apogee_cosmic_ruwe} shows our \cosmic results in blue, \textit{Gold Sample} in white, and \galah stars in the \textit{Gold Sample} in purple. The overlapping stars are well distributed among the \textit{Gold Sample}, and have similar range of visits. The final mass ratio for all samples in Figure \ref{fig:apogee_cosmic_ruwe} is defined as $q=M_2/M_1$, where $M_2<M_1$, to distinguish observable systems today. \apogee binaries have lower mass ratios, close to $q\sim0.01$, while the lowest \cosmic mass ratio is $\sim0.2$. The maximum period in the \textit{Gold Sample} is \app5000 days, while for \cosmic results is \app45,000 days. This difference in the mass ratios and orbital periods between the \apogee and \cosmic results is most likely because the \cosmic systems are a final configuration of a red giant and white dwarf, while the \apogee systems are not limited to a specific type of binary configuration. Based on Figure \ref{fig:apogee_cosmic_ruwe}, if any of our \lirich giants that were observed by \apogee were in binaries, we would only be able to detect systems with periods up to $s\approx 10,000$ days for mass ratios $> 10^{-1}$ and up to 50 days for mass ratios $< 10^{-2}$.

\subsection{Lithium production in AGB Stellar Models} \label{sec:models_results}


\begin{figure}
    \centering
    \includegraphics[width=\linewidth]{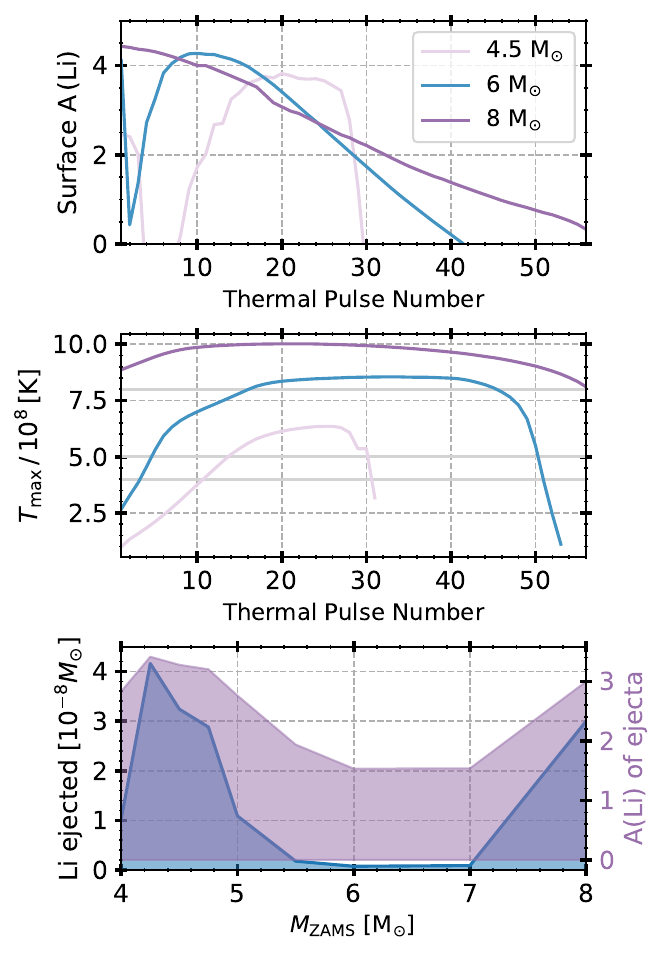}
    \caption{AGB models from \citet{karakas2014helium} and \citet{karakas2016stellar} with initial masses of 4.5, 6 and $8 \rm \; M_\odot$. The initial composition of all three models are approximately solar metallicity ($Z_i=0.014$) with canonical initial He mass fractions ($Y_i=0.28$). \textit{Top}: absolute surface \ali of the AGB as a function of thermal pulse number. \textit{Middle}: maximum temperatures reached at the base of the convective envelope during pulse cycle. \textit{Bottom}: total mass of Li ejected in the winds of models with initial masses $4 - 8 \rm \; M_\odot$ show on the left axis, and \ali of the ejecta as a function of initial AGB masses shown on the right axis.}
    \label{fig:monash_agb_genproperties}
\end{figure} 

In addition to the \cosmic simulations, we also consider the models from \citet{karakas2014helium,karakas2016stellar}. While \cosmic follows the dynamics of stellar systems, the Monash models allow us to follow the nucleosynthesis reactions that occur inside their interiors. Thus, we can probe their ability to produce lithium and the slow neutron capture elements. Given that the \lirich giants in \galah cover a broad metallicity distribution from $-1.8$ dex to $0.4$ dex, we consider how the initial mass of the star impacts nucleosynthesis processes at a token solar metallicity ($Z_{\rm i}=0.014$). The temperatures required for HBB during the TP--AGB are achieved in their $Z_{\rm i}=0.014$ models with $M_{\rm ZAMS} \geq 4.25\,$\solmass; thus, we consider their collection of $4-8$ \solmass models in the following analysis. 

In Figure \ref{fig:monash_agb_genproperties}, we show some of the key features of the AGB models. The upper panel demonstrates the transformation of surface \ali with each progressive thermal pulse. The central panel shows the temperatures at the base of the convective envelope; for burning to occur, temperatures need to exceed $\mathrm{\sim40 \; MK}$ for pp--reactions, $\mathrm{\sim50 \; MK}$ for CN reactions, and $\mathrm{\sim80 \; MK}$ for the full CNO cycle to operate (and the Mg--Al and Ne--Na chains). Note that these temperature thresholds are greater than those required for MS hydrogen burning, given the lower density profile at which HBB occurs. As discussed in Section \ref{sec:intro}, the production of $^{7}$Li nuclei occurs with the second chain of the proton--proton cycle (\textit{pp--}2). The frequency at which each hydrogen burning chain occurs depends sensitively on the temperature in the given region (see e.g., Section 18.5.1 of \citealt{kippenhahn2012stellar}). The \textit{pp--}2 reactions dominate for mild temperatures on the lower edge of the hydrogen burning threshold. With further increases in temperature, the frequency of \textit{pp--}3 reactions take over \textit{pp--}2, before CNO reactions dominate over all proton--proton chains.

The lower panel in Figure \ref{fig:monash_agb_genproperties} shows the net Li in solar masses in the mass expelled from each AGB model over its lifetime (left axis), and the total \ali of the eject (right axis). Peak Li in the ejecta occurs for the $4.25 \rm \; M_\odot$ model, where the state properties and abundances correspond to the appropriate temperatures for mild HBB reactions, as well as strong quantities of $^{3}$He fuel mixed into the envelope with the prior FDU episode. Only 5.7\% (59) of our \cosmic systems have an AGB with masses between $4.25-4.75$ \solmass. However, all models in the range $M_{\rm ZAMS} = 4.25 - 8$ contain at least $A(\rm Li) \geq 1.5$ in their expelled mass.




\subsection{Surface Abundance Changes After Mass Transfer} 
\label{sec:cosmic_models}

If we consider the \cosmic results from Section \ref{sec:cosmic_results}, the most likely mass transfer scenarios involving an intermediate--mass AGB and a lower mass companion were those that occurred on the main--sequence. This means that by the time we observe the red giant companion, the transferred material has been diluted by the expansion of the convective envelope into previous regions of hydrogen burning. For example, our 1 \solmass model has a 0.02 \solmass convective envelope on the main--sequence; this expands to a maximum of 0.7 \solmass during FDU, then retreats to $0.1-0.3$ \solmass during the RGB. 

\begin{figure}
    \centering
    \includegraphics[width=\linewidth]{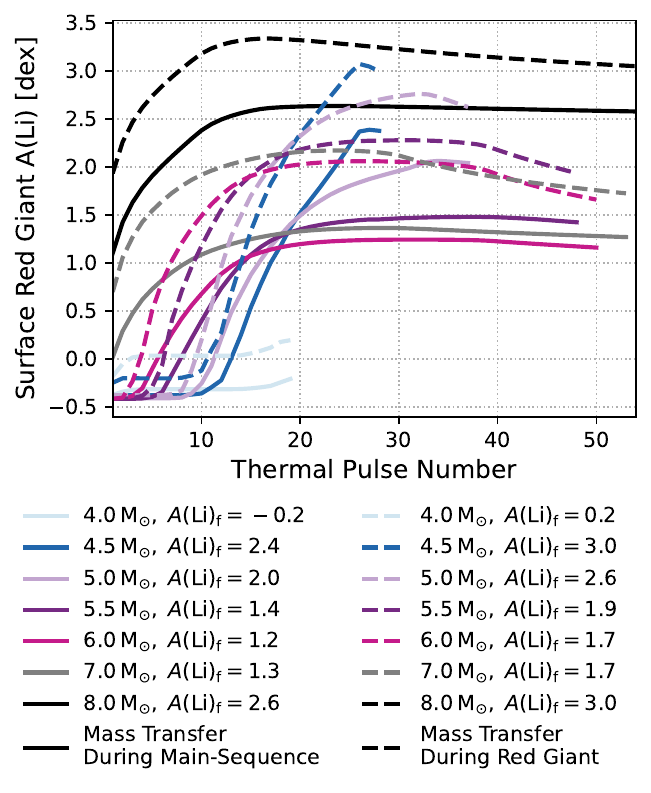}
    \caption{Surface \ali in a red giant after mass transfer from an AGB companion. Models are shown for varying AGB masses from $4-8$ \solmass. Solid lines model mass transfer during the red giant's main--sequence phase, and dashed lines show models when mass transfer occurs after the first dredge--up phase. Legend shows the final \ali post mass transfer for each model. Models have $Z_{\rm i} = 0.014$.}
    \label{fig:li_masses}
\end{figure}

Thus, to simulate the process of ‘main--sequence mass transfer’ with our models, we take the surface abundances (in mass fraction) of the intermediate mass AGB models and mix a percentage of this material (in \solmass) with a 0.7 \solmass envelope of post FDU abundances from a 1 \solmass Monash model. We mix the AGB abundances into the RGB envelope after each thermal pulse (using Equation \ref{eq:formula}), where we assume 5\% of the mass lost from the AGB is accreted onto the companion. We chose 5\% mass transfer efficiency as this resulted in a final lithium abundance consistent with the average \ali of the \galah sample. The resultant surface lithium abundance, \ali, on the companion is shown in Figure \ref{fig:li_masses}. 

Figure \ref{fig:li_masses} demonstrates that the greatest lithium pollution occurs due to donation from the 4.5, 5 and $8\rm \; M_\odot$ AGB models. Although \textit{pp}--2 reactions tend to dominate in the lower mass range (e.g., the 4.5 and $5\rm \; M_\odot$ models), the $8\rm \; M_\odot$ model experiences approximately two dozen more thermal pulses than the lower masses (and thus opportunities for more \textit{pp}--2 reactions to occur). This is demonstrated in the upper panel of Figure \ref{fig:monash_agb_genproperties}. Final $^7\rm Li$ quantities (surface mass fraction, net yield, and total mass expelled) for the intermediate mass models are presented in Table \ref{tb:lithium_finals}.

We also consider the less likely scenario from \cosmic, where the mass from the AGB is transferred during the red giants' giant phase of evolution. The main difference between this scenario and the former is the magnitude in mass of the giants’ convective envelope. The dashed lines in Figure \ref{fig:li_masses} show the mixed surface abundances with a $0.1\rm \; M_\odot$ convective envelope, typical of a red giant in \cosmic. Unlike the main--sequence mass transfer scenario, the red giant becomes \lirich for all but one AGB model. The $4\rm \; M_\odot$ model does not produce a \lirich giant given that it does not undergo HBB.

\begin{deluxetable}{c|c|c|c}[t!]
\tabletypesize{\footnotesize}
    \tablewidth{0pt}
    \tablecaption{Final $^7\rm Li$ values for the intermediate mass Monash models, where ${\rm X}(^7\rm Li)_f$ is the final surface mass fraction of the AGB. The net $^7\rm Li$ yield represents the integrated mass expelled from the model over its lifetime in comparison to initial abundances. The final column presents the net Li expelled from the model, in \solmass. \label{tb:lithium_finals}}
    \tablehead{\multicolumn{1}{c|}{AGB Mass} & \multicolumn{1}{c|}{${\rm X}(^7\rm Li)_f$} &\multicolumn{1}{c|}{
    Yield} & \multicolumn{1}{c|}{$M_{\rm tot}$ expelled} \\
    \multicolumn{1}{c|}{$\mathrm{[M_\odot]}$} & \multicolumn{1}{c|}{} & \multicolumn{1}{c|}{$\mathrm{[M_\odot]}$} & \multicolumn{1}{c|}{$\mathrm{[M_\odot]}$}
    }
    \startdata
    $4.25$ & 1.23E--08  & 1.09E--08 & 4.18E--08 \\ 
    $4.50$ & 8.91E--09  & --6.70E--10 & 3.24E--08 \\  
    $4.75$ & 7.42E--09  & --6.53E--09 & 2.88E--08 \\
    $5.00$ & 2.65E--09 & --2.66E--08 & 1.09E--08\\
    $5.50$ & 3.91E--10 & --4.00E--08& 1.80E--09\\
    $6.00$ & 1.52E--10  & --4.54E--08& 7.71E--10\\
    $7.00$ & 1.51E--10  & --5.39E--08& 9.11E--10\\
    $8.00$ & 4.32E--09  & --3.31E--08 & 3.00E--08\\
    \enddata
    \tablecomments{Positive total yields indicate that a given species is net produced over the models' lifetime, and negative yields indicate net destruction of a given species.}
\end{deluxetable}

\begin{figure*}
    \centering
    \includegraphics[width=\linewidth]{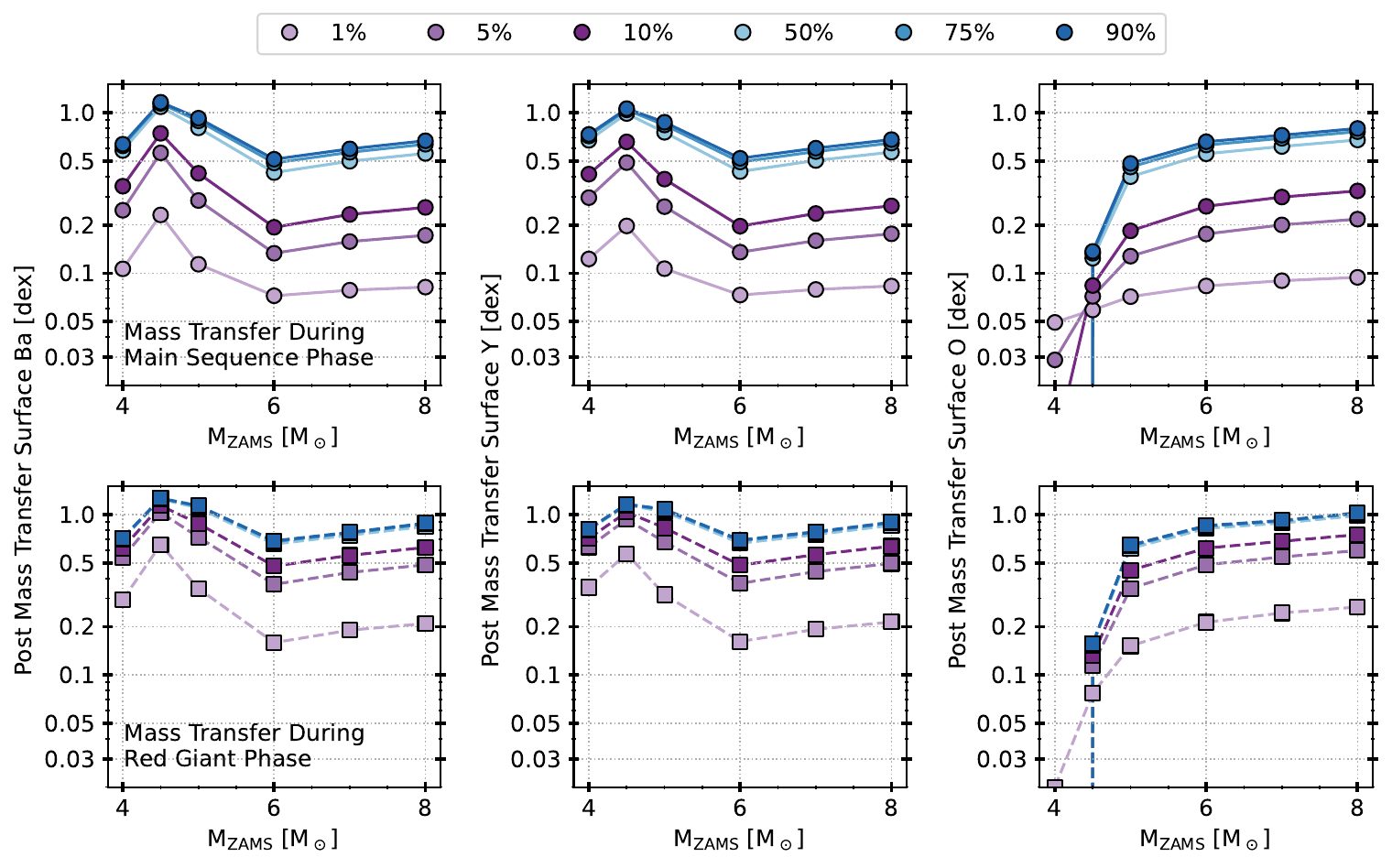}
    \caption{Changes in surface Ba (left), Y (middle), and O (right), post mass transfer in a 1 \solmass RGB for different mass transfer efficiencies as a function of donor AGB mass. The top panels show the abundance changes when mass transfer occurs on the main--sequence, and the bottom panels show abundance changes when \mt occurs on the red giant branch. The colours denote the varying \mt efficiencies as indicated in the legend.}
    \label{fig:eff_test}
\end{figure*}

\begin{figure*}
    \centering
    \includegraphics[width=\linewidth]{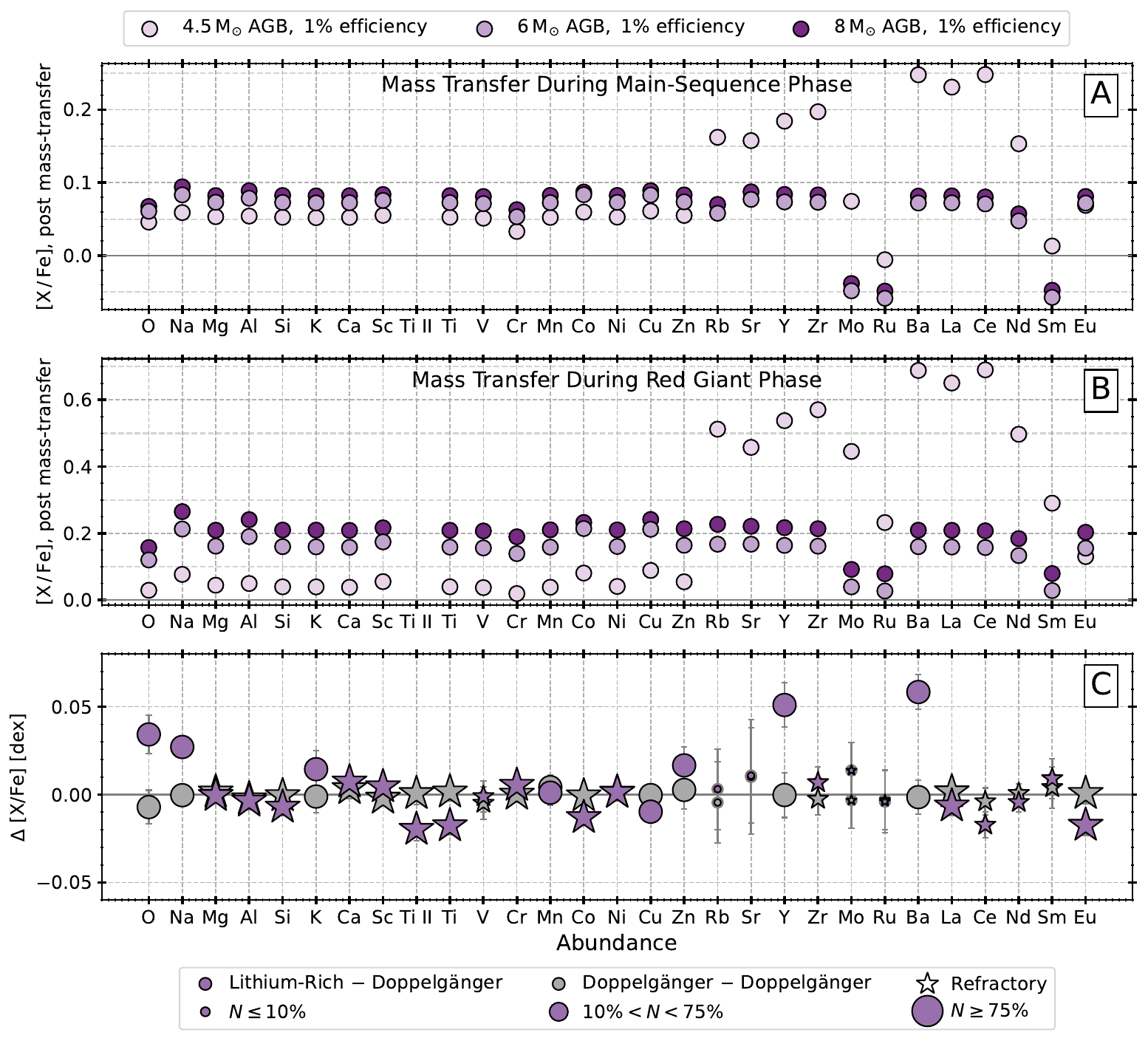}
    \caption{
    \textit{Top:} Surface abundance of a red giant after mass transfer from an AGB while the red giant is on the main--sequence. The six models shown are for varying AGB masses and mass transfer efficiency (see legend). The elements are ordered by their atomic number.
    \textit{Middle:} Same as top panel, but when the mass transfer occurs during the red giant phase. 
    \textit{Bottom:} Difference in abundance between a \lirich star and its \dg in \galah survey, as well as difference between two other \dgs for reference, where the difference is the average of all pairs. For each pair, the \dg is sampled 10 times and the difference is the median of 10 samples. The size of the markers represents the detection percentage, and the star marker denotes the refractory elements. The elements are ordered by their atomic number.}
    \label{fig:monash}
\end{figure*}

In addition to modeling Li, we use stellar models to investigate the surface abundance changes for other elements after mass transfer. However, we first test the effects of \mt efficiencies on resulting surface abundances for a range of AGB donor masses. In Figure \ref{fig:eff_test}, we test six \mt efficiencies from $1-90\%$ across AGB masses $4-8 \; \rm M_\odot$. We test the surface abundance changes in Ba, Y, and O, as these three elements showed the most significant changes between \lirich stars and their \dgs in \galah data in \cite{sayeed_2024}; $\rm \Delta(Ba) \approx 0.06$, $\rm \Delta(Y) \approx 0.05$, and $\rm \Delta(O) \approx 0.04$.  Moreover, we investigate changes in these elements when \mt occurs on the main--sequence (top panel in Figure \ref{fig:eff_test}) and on the RGB (bottom panel in Figure \ref{fig:eff_test}). Figure \ref{fig:eff_test} shows that surface abundance changes in Ba, Y, and O are most consistent with 1\% \mt efficiency across AGB masses. Surface changes after \mt occurs on the main--sequence are more similar to observed abundances as compared to when \mt occurs on the RGB. Based on Figure \ref{fig:eff_test}, we use 1\% \mt efficiency in our analysis with stellar models.

In Figure \ref{fig:monash}, we show the resultant changes in surface abundance after mass transfer for a broader spectrum of chemical species beyond lithium (Figure \ref{fig:li_masses}) and Ba or Y (Figure \ref{fig:eff_test}) as a function of atomic number, $Z$. We use 1\% \mt efficiency, but experiment with three AGB donor masses, 4.5, 6, and $\mathrm{8 \; M_\odot}$. These masses were selected based on their abundance yields in Figure \ref{fig:eff_test}; the $4.5 \; \rm M_\odot$ model shows the highest enhancements for all three elements (Ba, Y, and O), while the $6 \; \rm M_\odot$ model shows the lowest enhancements, and the $8 \; \rm M_\odot$ model was chosen as an edge case.

Figure \ref{fig:monash}A represents the change in surface abundance in a $\mathrm{1 \; M_\odot}$ red giant if mass transfer occurred during the main--sequence phase, while Figure \ref{fig:monash}B shows the resulting surface abundance changes if mass transfer occurred after FDU in the red giant. Interestingly, the $\mathrm{4.5 \; M_\odot}$ model shows larger enhancements for $Z=31-62$ range (Rb to Sm), but the $8 \; \rm M_\odot$ model shows larger yields  $Z<31$ and $Z>62$. The largest enhancements in both panels occurs for $4.5 \; \rm M_\odot$ model, while the other two models (6 and $8 \; \rm M_\odot$) show similar yields across abundances. The highest enhancements are found in the \sprocess element family, namely Sm, Rb, Sr, Y, Zr, Mo, Ba, La, Nd, irrespective of the specific model. 

Figure \ref{fig:monash}B shows mass transfer post FDU, when the red giant has left the main--sequence, which results in higher enhancement in all abundances as compared to results in \ref{fig:monash}A. Similar to \ref{fig:monash}A, the $\mathrm{4.5 \; M_\odot}$ model shows the strongest enhancements for elements $Z=31-62$ at a given efficiency, whereas the $\mathrm{4.5 \; M_\odot}$ model again shows higher quantities of the \sprocess elements. 

\subsection{Comparing \galah Observations to Stellar Models} \label{sec:galah_models}

We compare the Monash models shown in Panels A and B in Figure \ref{fig:monash} to \galah abundances from the S24 \lirich sample in Figure \ref{fig:monash}C. Panel C shows the difference in each abundance between a \lirich and a \dg in purple, the difference between two other \dgs in grey, and refractory elements in star marker. We randomly select \dgs from 50 closest \dgs, where we ensure that the difference in signal--to--noise between the \lirich and its potential \dgs is below 50, and the selected \dg has a good quality flag in \galah. To ensure high fidelity, we randomly select a \dg 10 times, and use the median of all samples as the difference between a \lirich star and its \dg. We then take the average difference between all pairs for each element, where the marker size denotes the detection efficiency: the number of pairs with available measurements out of the maximum number of pairs with available measurements, $N=812$. Furthermore, \galah DR3 cautions against using measurements for the following elements as they might be affected by blending issues: Co, La, Mo, Nd, Rb, Ru, Sm, Sr, V, and Zr. Therefore, for these elements, we remove any \lirich stars and \dgs with measurements [X/Fe] $> 0.3$ dex, as suggested by \cite{galah_sven_2021}. We also remove stars with problematic measurements for Ti II in \galah DR3 due to empirically fixed microturbulence velocity that overestimate Ti II values for some red clump stars \citep[for a discussion see Section 6.5.1 by][]{galah_survey}. When fitting microturbulence during the parameter estimation, these velocities would differ beyond $0.25\,\mathrm{km\,s^{-1}}$ and thus lead to lower Ti II abundances \citep{galah_sven_2021,galah_dr4}.

Despite using the same sample of \lirich giants, there are some differences between our Figure \ref{fig:monash}C and Figure 12 in S24. Unlike S24, we added an additional SNR cut and removed suspicious measurements for 10 elements given the recommendation by \cite{galah_sven_2021}. While elements Ba, O, and Y are similarly enhanced, K, V and Zn now show a negligible difference between the \lirich and reference sample. In addition, although S24 did not show Mo, Sr, Ru, and Rb due to a smaller sample size, we show them in Figure \ref{fig:monash}C for completeness. 

In Figure \ref{fig:monash}C, the \sprocess elements Ba and Y are enhanced in \lirich stars as compared to their \dgs, while the remaining \sprocess elements show negligible difference between \lirich star and their \dgs. Compared to the models shown in Panels A and B in Figure \ref{fig:monash}, the \sprocess elements in \lirich giants in the \galah sample are significantly less enhanced. In both \mt scenarios, the $6 \rm \; M_\odot$ model produces yields that are the most consistent with \galah observations. \galah observations show that Ba and Y measurements for \lirich giants as compared to \dgs are 0.058 dex and 0.051 dex, respectively. In the scenario when \mt occurs on the main--sequence (Panel A), the $6 \rm \; M_\odot$ model results in surface Ba and Y of 0.072 dex and 0.073 dex, respectively. On a linear scale, (ie. $10^{\rm [X/Fe]_{model}}/10^{\rm [X/Fe]_{GALAH}}$), there is a $95-97\%$ agreement between the observed Ba value and the values from 6 and $8 \rm \; M_\odot$, and a $93-95\%$ agreement for Y. We did not compare the yields for the seven other \sprocess elements (La, Mo, Nd, Rb, Sm, Sr, Zr) as the uncertainties for these measurements are large in \galah DR3 which make them unreliable for comparison.

The deviation between the observed and modeled abundances could be due to multiple reasons. Firstly, our observations in Figure \ref{fig:monash}C show the average of a population rather than an individual star shown in Figure \ref{fig:monash}A/B. Secondly, the stellar models in Figure \ref{fig:monash}A/B are shown for a red giant with solar mass and solar metallicity, and an AGB donor of 4.5, 6, and 8 \solmass. Our \lirich red giants possess a range of masses and metallicities, and cannot be represented fully by one model. Lastly, these discrepancies could in fact suggest that this channel is not a viable method for Li--enrichment of red giants. For instance, \cite{Rybizki2017} used chemical evolutionary model \textit{Chempy} to model the abundances in the Sun, Arcturus, and the local, present--day ISM traced by B--stars. They found discrepancies between observed and predicted abundances for certain elements irrespective of underlying hyper--parameters and yield sets, and suggest that these discrepancies could be an indication of missing nucleosynthetic channels.

Despite there being a difference in  mass and metallicity between the stellar models shown in Figure \ref{fig:monash}A/B and the average of the \lirich population shown in Figure \ref{fig:monash}C, two combinations of initial mass and mass transfer efficiency scenarios produce similar amounts of \sprocess abundances in comparison to \galah observations, namely the 6 \solmass and 8 \solmass AGB with 1\% mass transfer efficiency. Based on our stellar models, we conclude that enhancements in \sprocess elements are feasible in a \lirich giant with an AGB companion.

In addition, we see non--negligible enhancements between \lirich stars and their \dgs for Na, O, K, and Zn in \galah observations in Figure \ref{fig:monash}C which are not similarly enhanced in the models. In Panel C, the 3rd largest enhancement is seen in O. The \galah enhancement in O between \lirich stars and their \dgs is 0.034 dex, while the stellar model yields for the main--sequence \mt scenario are 0.045, 0.061, and 0.067 dex, for the 4.5, 6, and $8 \rm \; M_\odot$ models, respectively. On a linear scale, there is a $92-97\%$ agreement in O yields between \galah observations and stellar models. The Na, K, and Zn enhancements are 0.027 dex, 0.014 dex, and 0.017 dex, respectively. On a linear scale, there is a $85-93\%$ agreement between \galah observations and stellar models for these three elements.

There could be multiple reasons for the small discrepancy. Firstly, the cause could be due to systematic uncertainties in the measurements of these elements. Using the same sample and method, we re--ran the analysis with \galah DR4 data \citep{galah_dr4}, and found that \lirich stars were enhanced to non--negligible amounts in Ca, Na, and Zn as compared to their \dgs, but still dissimilar to AGB models in Figure \ref{fig:monash}A/B. However, it is important to note that none of these elements are produced in our current AGB models, and so different models (ie. massive stars models), or initial conditions are needed for accurate comparison. For instance, Na and Zn are produced during supernovae explosions (e.g., supernovae Type Ia and Type II; \citealt{Kobayashi2020, arcones2023origin, Matsuno2025}) which we do not follow. Nevertheless, if the differences in Na, O, K, and Zn in Figure \ref{fig:monash}C are astrophysical, these elements may indicate that there are multiple mechanisms responsible for \lirich giants. For instance, both Na and O have been theorized to form in AGB stars with HBB, but the yields are strongly dependent on reaction rates, and the majority of Na and O is made elsewhere, such as SNe Ia and SNe II sites \citep[e.g.,][]{Herwig2000}.

\section{Discussion} \label{sec:discuss} 
We combine our results from \cosmic simulations (Section \ref{sec:cosmic_results}) and stellar models (Sections \ref{sec:models_results} -- \ref{sec:galah_models}) to make predictions of possible pathways for Li--enrichment of red giants. We describe our predictions in Section \ref{sec:predictions}, the limitations and caveats of our results in Section \ref{sec:limits}, and the implications of our results with the upcoming \gaia DR4 data in Section \ref{sec:gaia}.

\subsection{Predictions for Li--enrichment} \label{sec:predictions}
\begin{figure}[t!]
    \centering
    \includegraphics[width=\linewidth]{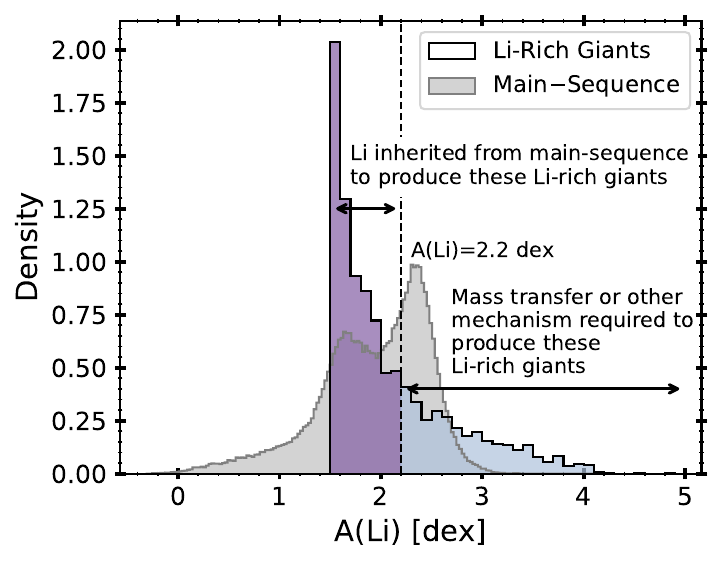}
    \caption{Distribution of lithium abundance for main--sequence (grey) and \lirich red giants in \galah. The vertical dashed line indicates $\rm A(Li)=2.2 \; dex$ for reference. }
    \label{fig:pred}
\end{figure}

Our \cosmic simulations find that \mt from an AGB companion to the RGB occurs during the red giant's main--sequence phase 98\% of the time. Given the high probability of this scenario, we use \galah data to test the likelihood of this scenario. The main--sequence lithium measurements from \galah combined with expectation for first dredge--up depletion shows that main--sequence stars with $A\mathrm{(Li)=1.5-2.2 \; dex}$ can become \lirich giants without external mechanisms. During FDU, we expect lithium to be depleted by $\sim 1 \; \rm dex$ as suggested by previous studies \citep[e.g.,][]{Lind2009, Karakas2014_review, Charbonnel2020}. As a result, a main--sequence star must have $\rm A(Li)\geq2.5$ dex in order to remain \lirich as a red giant after FDU (ie. $\rm A(Li) \geq 1.5 \; dex$). Figure \ref{fig:pred} illustrates this idea by comparing the lithium abundance distribution of main--sequence stars in \galah to \lirich giants. In \galah, 9.2\% of main--sequence\footnote{\teff $= 3000-6500$ K, \logg $=4.0-5.2$ dex, \texttt{flag\_sp <= 1}, \texttt{flag\_fe\_h = 0}, $E(B-V)<0.33$.} stars have $\rm A(Li) = 2.5-3.2 \; dex$. Therefore, it is possible for Li--enriched main--sequence stars with  $\rm A(Li)=2.5-3.2 \; dex$ to become \lirich giants (ie. $\rm A(Li)=1.5-2.2$ dex). 

We further emphasize that although 9.2\% of main--sequence stars enhanced in lithium have the potential to produce \lirich giants (i.e., red giants with $\rm A(Li)\geq1.5$ dex), the depletion level during FDU is highly variable. We adopt a depletion level of 1 dex for our calculations in this work \citep[e.g.,][]{Lind2009, Karakas2014_review, Charbonnel2020}; however, the depletion level can be as large as 2 dex \citep[e.g.,][]{iben1967_vi, iben1967_vii}. Moreover, we find only 0.04\% of main--sequence stars have $A\rm{(Li)}\geq3.2 \; dex$; this fraction is too small to explain the 32\% of \lirich giants with $A\rm{(Li)}\geq2.2 \; dex$. Therefore, a red giant with $\rm A(Li) \geq 2.2 \; dex$ requires some additional pathway to achieve this enrichment.

Our analysis with \cosmic and AGB models enables us to constrain the most probable parameters of binaries that may underlie \lirich giants, in particular for those with $\rm A(Li) \geq 2.2$ dex. Based on our results from \cosmic simulations, AGB stellar models, and chemical abundances from \galah, we propose that these objects are most likely created when mass is transferred from an intermediate--mass AGB onto a red giant during its main--sequence phase via a \ce. Mass transfer during red giant phase is less likely due to stellar evolutionary timescales. Moreover, \cosmic simulations show that mass transfer during the red giant phase only occurred in three instances, and when the AGB mass was between $\rm 1.3-1.6 \; M_\odot$ which is not sufficiently massive enough to produce lithium. The remaining scenario of mass transfer during the main--sequence phase is not only more likely, but also typically involves a Li--producing, intermediate--mass AGB. The possibility of no CE in our scenario is unlikely given that \cosmic analysis shows that systems with RLOF produced a CE for 82\% of systems.

Additionally, stellar models support the possibility of a red giant becoming \lirich with and without high Li content initially. Figure \ref{fig:calcs} shows that Li--enrichment is possible even with small envelope and accreted masses. For instance, assuming an AGB companion with $\rm A(Li) =3.2 \; dex$ and a red giant with $\rm A(Li)= -0.4 \; dex$, the red giant could become \lirich after accreting as little as $\rm 0.1 \; M_\odot$ from the AGB. In addition, Figure \ref{fig:li_masses} shows a red giant can achieve $\rm A(Li)=2.6 \; dex$ after mass transfer from a $\rm 8 \; M_\odot$ AGB even if mass transfer occurred on the main--sequence.

In summary, \lirich giants with $\rm A(Li) =1.5-2.2 \; dex$ could be produced even without mass transfer from an intermediate--mass AGB, while \lirich giants with $\rm A(Li)\gtrsim 2.2 \; dex$ could only be produced with mass transfer, under the assumption of binarity. This is further supported by the fact that in our sample of \galah \lirich giants, the mean Ba abundance, a product of intermediate--mass AGB stars, is higher in giants with $\rm A(Li)\geq 2.2\;dex$, $\mu_{\rm Ba}=0.31\pm0.01$ than those below this threshold, $\mu_{\rm Ba}=0.19\pm0.01$. We also use stellar models to test the abundance characteristics of mass transfer from an AGB companion. Mass transfer from a $8 \; \rm M_\odot$ AGB can provide enhancements in \sprocess elements similar to those seen in \galah data for \lirich giants. However, it is important to note that enhancements seen in Figure \ref{fig:monash} are for a specific AGB mass, chosen red giant envelope mass, and red giant mass. These enhancements are less pronounced for varying scenarios, such as for a higher efficiency or lower mass AGB.

\subsection{Assumptions \& Simplifications}
\label{sec:assumptions}

We note that the resulting architectures change significantly based on the choice of prior for the primary's mass. For instance, the number of systems with an intermediate--mass AGB is only 4\% if we choose a Gaussian distribution centered on 1.5 \solmass as compared to 35\% if we choose a uniform prior between $0.5-10$ \solmass. Similarly, the number of systems that undergo RLOF changes to 32\% rather than the current 55\%. We chose to use a uniform distribution to sample from for the primary's mass to ensure a large sample size for systems that undergo RLOF and end up with intermediate--mass AGB companions. While the relative numbers of systems may vary, our main conclusions are consistent between the two choices of priors. For instance, \mt occurs on the main--sequence for 98\% of systems given a Gaussian prior, as compared to almost 100\% of systems given a uniform prior.

Furthermore, our metallicity prior is centered on the mean metallicity of \lirich giants in \galah, $\mu_{[\rm Fe/H]}=-0.34 \rm \; dex$, or $\mu_{Z}=0.5 \; Z_\odot$. However, metallicity significantly impacts the radial evolution of a star due to opacity, thereby affecting the binary interactions with a companion \citep[e.g.,][]{Klencki2020}. Stars at lower metallicity are more compact and hotter, and have a higher supply of hydrogen, causing longer lifetimes. \cite{Klencki2020} found that at lower metallicity, the primary (or donor) star is more evolved at the stage of RLOF for massive donors. In addition, our \cosmic simulations found that the onset time of RLOF is $100-200 \rm \; Myr$ earlier for stars at 10\% solar metallicity as compared to stars at solar metallicity. Therefore, a change in metallicity prior would largely impact the lifetimes of evolutionary stages, and have minimal effect on the final results. For instance, since we do not restrict the relative time at which the star goes through a specific evolutionary phase, the change in metallicity would have negligible effect on our final conclusions. In addition, our stellar models assume $Z_\odot=0.014$ in comparison with \cosmic which assumes $Z_\odot=0.019$. We find a negligible difference in the internal physics in Monash stellar models if we assume a different solar metallicity. This is consistent with \cite{Karakas2014_review} who found no difference in stellar structure between $Z_\odot = 0.02$ and $Z_\odot = 0.014$ for a $\rm 3\; M_\odot$ AGB star.

In addition, we assume a mass transfer efficiency ($\beta$) of 1\% in our stellar models, as in 1\% of the mass is transferred to the red giant from the primary. Observationally, the mass transfer efficiency in binary systems is not constrained; therefore, binary systems with varying configuration and binary parameters cannot be all explained by one choice of $\beta$ \citep[e.g.,][]{demink2007}. In fact, some studies find decreased mass transfer efficiency for increased orbital periods \citep[e.g.,][]{demink2007, sen2022}. We explored the effect of mass transfer efficiency on abundance yields in Figure \ref{fig:eff_test}, where we choose $\beta = 0.01, 0.05, 0.1, 0.5, 0.75, 0.9$ assuming all else constant (ie. $M_{\rm AGB}$, $M_{\rm RGB}$, evolutionary state of RGB during mass transfer). We find that the model where $\beta = 0.01$ was the best match to our observations of \lirich giants in \galah (Panel C in Figure \ref{fig:monash}) when comparing lithium and \sprocess elements. Given that the lower $\beta$ value -- although uncommon \citep[e.g.,][]{Nelson2001} -- is a better match to our observations, this result suggests that direct \mt via RLOF is not the preferred method of \mt. Inefficient \mt, as is the case with wind RLOF or Bondi--Hoyle--Lyttleton accretion, would produce a lower \mt efficiency. Since \cosmic does not account for wind RLOF, probing binary architectures in this \mt scenario is out of the scope of this paper.

\subsection{Limitations and Caveats}
\label{sec:limits}
In this work we test binary systems only focusing on the \mt route. However, it is plausible that \lirich objects are formed via other means. For instance, triple systems have been offered as a possible explanation for other similarly rare phenomenon \cite[e.g.,][]{Dodd2024}. Another explanation is that these objects are remnants of stellar mergers \citep[e.g.,][]{Zhang2013}. By performing binary population synthesis simulations, \cite{Zhang2020} found that \lirich stars on the red clump can be produced as a result of a merger between a low--mass helium white--dwarf ($0.35-0.40$ \solmass) and a red giant. In Figure \ref{fig:hrd}, we find that red clump stars tend to have higher \ali than stars across the red giant branch. This is similar to results from S24 that found red clump stars tend to have higher Li abundance. Therefore, it is possible that \lirich stars on the red clump with $\rm A(Li) \gtrsim2.2 \; dex$ are a product of stellar mergers. Evidence of stellar mergers can be further investigated with asteroseismology. \cite{Rui2024} investigated post merger products between helium white--dwarfs and main--sequence stars; their models showed Li--enrichment immediately after the merger due to dredge--up from a hydrogen burning event, but they noted their models were artificial and unreliable. Although they considered He--flash as a formation mechanism for Li--enrichment in giants \citep[e.g.,][]{kumar_2020, Mallick_2023}, their models did not produce \lirich giants (ie. $\rm A(Li) \geq 1.5 \; dex$). 

Moreover, our \cosmic results indicate that Li--enrichment occurs before the red giant phase. In \cosmic, there were no systems where \mt took place during the red giant phase where the AGB companion could have undergone HBB (ie. $M=4-8 \rm \; M_\odot$). Since HBB only occurs in intermediate--mass AGB stars, mass transfer during the red giant phase is not likely. Furthermore, the timescale of the red giant phase are shorter than the main--sequence phase for intermediate to high mass stars which makes \mt less likely during the red giant phase. An outstanding possibility is tidal spin--up from a binary companion \citep{casey_2019}, but this remains to be quantitatively tested. 

A final scenario in which Li--enrichment could occur is with Bondi--Hoyle--Lyttleton accretion. However, we do not explore this scenario as a primary mechanism for Li--enrichment as it is not implemented in \cosmic; however we briefly explore this \mt scenario below. This accretion scenario would occur if the separation is too large or if the dust formation region is smaller than the Roche--Lobe of the AGB star \citep[e.g.,][]{chen2020}. Cool stars with $L/M > 10^3 \; L_\odot/M_\odot$ have dust driven winds where pulsations extend the photosphere and enable formation of dust. Stellar winds from AGB stars can reach 10 times the stellar radius where the typical radii for AGB stars is $\rm 1 \; AU$ \citep{Hofner2018}, and terminal wind velocity is $5-20$ \kms \citep{chen2017, vassiliadis1993evolution, Groenewegen2007}. Our \cosmic simulations indicate systems with no RLOF are found at a mean final separation of $\rm \mu=13 \; AU$ and $\rm \sigma=6 \; AU$. The best stellar model that is consistent with our data is $6\rm \; M_\odot$ based on Figure \ref{fig:monash}. Assuming a 1\% \mt efficiency, 0.06 \solmass is transferred to the $1 \rm \; M_\odot$ RGB during its main--sequence phase. If we assume a mass loss rate of $10^{-8} \rm \; M_\odot \; year^{-1}$ \citep{Hofner2018}, the RGB can gain 0.06 \solmass from the AGB in 6 Myr, which is consistent with the typical lifetime of an AGB system in our \cosmic simulations of \app 3 Myr.

\begin{figure}[t!]
    \centering
    \includegraphics[width=\linewidth]{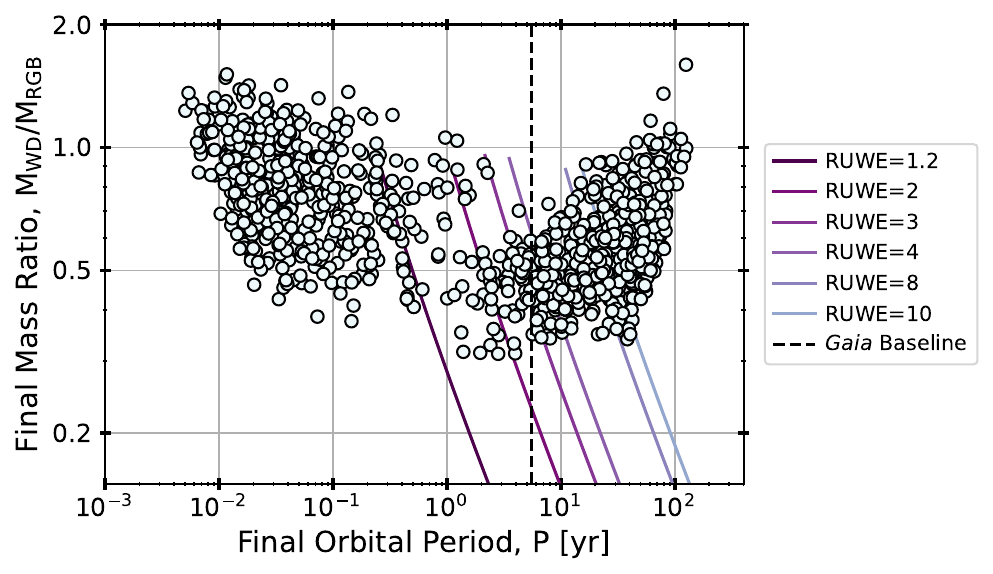}
    \caption{Final mass ratio and orbital period (in years) of \cosmic systems with simulated RUWE curves. The vertical black line indicates \gaia DR4 baseline of 66 months.}
    \label{fig:ruwe}
\end{figure}

\subsection{Observing Binaries with \gaia DR4} \label{sec:gaia}

The goal of this work is to probe binary architectures of \lirich giants -- under the assumption they host companions -- for future radial velocity follow--up. \gaia's RUWE values indicate photometric contamination from a companion at wide orbits ($\mathrm{\geq 2 \; AU}$). Most of the \galah \lirich giants have $\mathrm{RUWE \; < 1.4}$, which is the threshold for expected contamination from binaries. However, if the \lirich giants are on close orbits ($\lesssim 2$ AU), they would be insensitive to RUWE values. Fortunately, \gaia DR4 (scheduled for end of 2026) is expected to provide radial velocity measurements for all sources. If our matched \cosmic--\galah systems are accurate, then we can expect to test binarity of \lirich giants for systems shorter than \gaia observing span of at least 66 months. Figure \ref{fig:ruwe} shows simulated RUWE curves between RUWE $=1.2-9$ for a range of mass ratios and orbital periods, over--plotted on our \cosmic results for reference. These lines are computed by taking the sky position of all \galah \lirich giant stars, and their apparent magnitudes, in order to compute the approximate RUWE given some orbital period and a circular orbit. Since the RUWE will vary by sky position (all else being equal), the lines show the sky--averaged value. For fainter or more distant samples, these relationships are expected to vary. The vertical black line indicates the \gaia baseline of 66 months. While many systems fall below the \gaia observing baseline, the detectability of binaries is sensitive to multiple factors, such as baseline and number of visits, and does not guarantee observability \citep[e.g.,][]{Wallace2025}.

\section{Conclusions} \label{sec:conclusion}
In an effort to better understand the complex and likely diverse contributions to Li--enhancement of red giants, our study aimed to put constraints on one of the hypothesized pathways: mass transfer from an AGB companion to a red giant. We used the binary population synthesis code \cosmic to simulate possible architectures of \lirich giants under the assumption these objects become Li--enhanced due to binary interactions. We use observed properties of \galah giants to initialize input parameters into \cosmic, including age, mass and metallicity, until the system reaches the observed age of a \lirich giant in \galah. To test the role of AGB stars in the formation of \lirich giants, we require that the final system consists of a red giant and white--dwarf, and that the white--dwarf has undergone an AGB phase. We use stellar models to predict the changes in surface abundance of a red giant after mass transfer from an intermediate--mass AGB. Based on \cosmic simulations, AGB models, and \galah observations, our main conclusions are as follows:

\begin{enumerate}[i)]
    \item \lirich giants with $A\mathrm{(Li)=1.5-5.0}$ dex can be created with \mt from an intermediate--mass ($4.25-8$ \solmass) AGB with a CE when the giant is on the main--sequence. These correspond to 29\% of \cosmic systems of the total 1021 systems, at final mean separations of $3.5\pm0.4$ AU and final mass ratios of $0.5-5.0$; see  Figure \ref{fig:qi}. 
    \item Our stellar models show maximum Li in the ejecta of an AGB occurs for $M=4.25-4.75$ \solmass, which corresponds to only 5.7\% of our 1021 \cosmic systems. These systems are at mean separations of $5.1\pm1.1$ AU and mass ratios $0.5-1.1$.
    \item Based on \galah observations of \lirich giants, we suggest that it is possible for giants with $A\mathrm{(Li)=1.5-2.2}$ dex to become \lirich by inheriting Li from their main--sequence phase rather than an enrichment mechanism; however, an enrichment mechanism is required for a main--sequence star to become Li--enhanced. \cosmic results indicate that \mt is less likely during the red giant phase of the accretor.  
    \item By modeling the changes in surface abundances of a $1\rm \; M_\odot$ RGB after mass transfer from a 4.5, 6, and $8\rm \; M_\odot$ AGB at 1\% mass efficiency, we recovered similar enhancements for only two \sprocess elements, Ba and Y for both 6 and $8\rm \; M_\odot$ AGB model. The amplitude of these enhancements is comparable to that measured in the population of \lirich stars, first seen in \cite{sayeed_2024}. The difference in amplitude between the measurement from the data and expectation from theoretical stellar models could be because our models reproduce surface abundance for a single star with a specific metallicity and mass, while our \galah observations show the average of a population of stars with varying stellar properties.
    \item We compare \cosmic results with detected \apogee binaries, and conclude that if \galah \lirich giants were observed by \apogee, we could only detect systems up to $s\approx 10,000$ days as this is the upper limit of detected binaries in \apogee \citep{apw_2020}. 
\end{enumerate}

We constrain binary architectures of \lirich giants assuming binarity as a formation mechanism for these objects. However, other Li--enrichment channels not tested here remain possible routes of enrichment, including planet engulfment \citep[e.g.,][]{DelgadoMena2014, Martell2021}, and the He--flash \citep[e.g.,][]{kumar_2020, Zhang2021, Mallick_2023}. With \cosmic, we now have a view into the parameter space of \lirich giants binary systems. Upcoming \gaia DR4 data will provide radial velocity measurements for all sources which could reveal this binarity empirically. However, binary confirmation with \gaia DR4 is not guaranteed even for intrinsic binary systems due to factors that affect the detectability of the system, such as apparent magnitude, sky location, orbital period, the number of and spacing between visits, among other factors.  Therefore, dedicated radial velocity follow--up is also needed in this domain to extensively test binarity of \lirich giants. 

\section{Acknowledgments}
We are thankful to our anonymous referee for helpful comments that improved this manuscript. MS would like to acknowledge the support of the Natural Sciences and Engineering Research Council of Canada (NSERC). Nous remercions le Conseil de recherches en sciences naturelles et en génie du Canada (CRSNG) de son soutien. MS thanks the LSSTC Data Science Fellowship Program, which is funded by LSSTC, NSF Cybertraining Grant \#1829740, the Brinson Foundation, and the Moore Foundation; her participation in the program has benefitted this work. This work was supported by the Australian Research Council Centre of Excellence for All Sky Astrophysics in 3 Dimensions (ASTRO 3D), through project number CE170100013. SB acknowledges support from the Australian Research Council under grant number DE240100150. The Flatiron Institute is a division of the Simons Foundation.

\bibliography{references}{}
\bibliographystyle{aasjournal}

\appendix
\label{figuresappendix} 
\renewcommand{\thefigure}{A\arabic{figure}} 
\setcounter{figure}{0}

\section{Abundance Calculation} \label{sec:calculation}

In order to calculate a surface abundance post mass transfer, we first convert number fraction $N(i)$ -- which is the output for each species in our Monash code \citep{LattanzioThesis, Lattanzio86, Frost96, Karakas07} -- into mass fraction, $X(i)$:

\begin{equation}
    X(i) = N(i) \times M_i,
\end{equation}

where $M_i$ is the atomic mass for a given species $i$. We calculate the new mixed abundances in terms of mass fraction using equation \ref{eq:formula}, then convert the final mixed abundances ($X_\textrm{RGB,a.a.}$) back into number fraction by dividing through with the atomic mass $M_i$, 

\begin{equation}
    N(i) = \frac{X(i)}{M_i},
\end{equation}

and finally into absolute abundances, $A(i)$:

\begin{equation}
    A_\textrm{i} = \log_{10}(N_\textrm{i}/N_\textrm{H}) + 12
\end{equation}

where $N_\textrm{H}$ is the number fraction of hydrogen. To convert to [$i$/Fe] notation, calculate [$i$/H] first, 

\begin{equation}
    [i\textrm{/H]} = A_\textrm{i} - A_{\textrm{i},\odot},
\end{equation}

and new [Fe/H], 

\begin{equation}
    [\textrm{Fe/H}] = A_\textrm{Fe} - A_{\textrm{Fe},\odot},
\end{equation}

finally, 

\begin{equation}
    [i\textrm{/Fe]} = [i\textrm{/H]} - \textrm{[Fe/H] }
\end{equation}



\end{document}